\documentclass{jfm}

\usepackage{graphicx}
\usepackage{newtxtext}
\usepackage{newtxmath}
\usepackage{natbib}
\usepackage{hyperref}
\usepackage[ruled,linesnumbered]{algorithm2e}
\hypersetup{
	colorlinks = true,
	hidelinks,
	urlcolor   = black,
	citecolor  = black,
}

\newcommand{\RomanNumeralCaps}[1]
\linenumbers

\definecolor{dgreen}{rgb}{0.2,0.6,0.2}
\definecolor{pink}{rgb}{0.9,0,0.5}

\title{Vortex–magnetic competition and regime transitions in antiparallel flux tubes}
\shorttitle{Vortex–magnetic competition in flux tubes}
\shortauthor{W. Shen, R. Ostilla-Mónico and X. Zhu}

\author{Weiyu Shen\aff{1},
	Rodolfo Ostilla-Mónico\aff{2},
	\and Xiaojue Zhu\aff{1}
	\corresp{\email{zhux@mps.mpg.de}}}

\affiliation{
	\aff{1}Max Planck Institute for Solar System Research, 37077 Göttingen, Germany
	\aff{2}Departamento de Ingeniería Mecánica y Diseño Industrial, Escuela de Ingeniería, Universidad de Cádiz, 11519 Puerto Real, Spain}

\begin{document}
\maketitle
	
\begin{abstract}
Vortex–magnetic interactions shape magnetohydrodynamic (MHD) turbulence, influencing energy transfer in astrophysical, geophysical, and industrial systems. On the Sun, granular-scale vortex flows couple strongly with magnetic fields, channeling energy into the corona.
At high Reynolds numbers, vorticity and magnetic fields are nearly frozen into the charged fluid, and MHD flows emerge from the Lorentz force mediated interactions between coherent vortex structures in matter and the field. To probe this competition in a controlled setting, we revisit the canonical problem of two antiparallel flux tubes. By varying the magnetic flux threading each tube—and thus sweeping the interaction parameter $N_i$, which gauges Lorentz-to-inertial force balance—we uncover three distinct regimes: vortex-dominated joint reconnection, instability-triggered cascade, and Lorentz-induced vortex disruption.
At low $N_i$, classical vortex dynamics dominate, driving joint vortex–magnetic reconnection and amplifying magnetic energy via a dynamo effect.
At moderate $N_i$, the system oscillates between vorticity-driven attraction and magnetic damping, triggering instabilities and nonlinear interactions that spawn secondary filaments and drive an energy cascade.
At high $N_i$, Lorentz forces suppress vortex interactions, aligning the tubes axially while disrupting vortex cores and rapidly converting magnetic to kinetic energy.
These findings reveal how the inertial–Lorentz balance governs energy transfer and coherent structure formation in MHD turbulence, offering insight into vortex–magnetic coevolution in astrophysical plasmas.
\end{abstract}


\section{Introduction}\label{sec:intro}

Magnetohydrodynamics (MHD) governs a wide range of astrophysical, geophysical, and industrial phenomena, from solar flares \citep{ Sweet1969Mechanisms,Shibata2011Solar} and stellar dynamos \citep{Charbonneau2023Evolution,Warnecke2023Numerical} to fusion plasmas \citep{Ichimaru1993Nuclear} and planetary magnetospheres \citep{Johnson2014Kelvin}. At the core of these processes lies the intricate interplay between vortices and magnetic fields, which fundamentally shapes energy transfer, turbulence development, and the evolution of MHD structures \citep{Davidson2013Turbulence}.

\begin{figure}
	\centering
	\includegraphics[width=0.75\linewidth]{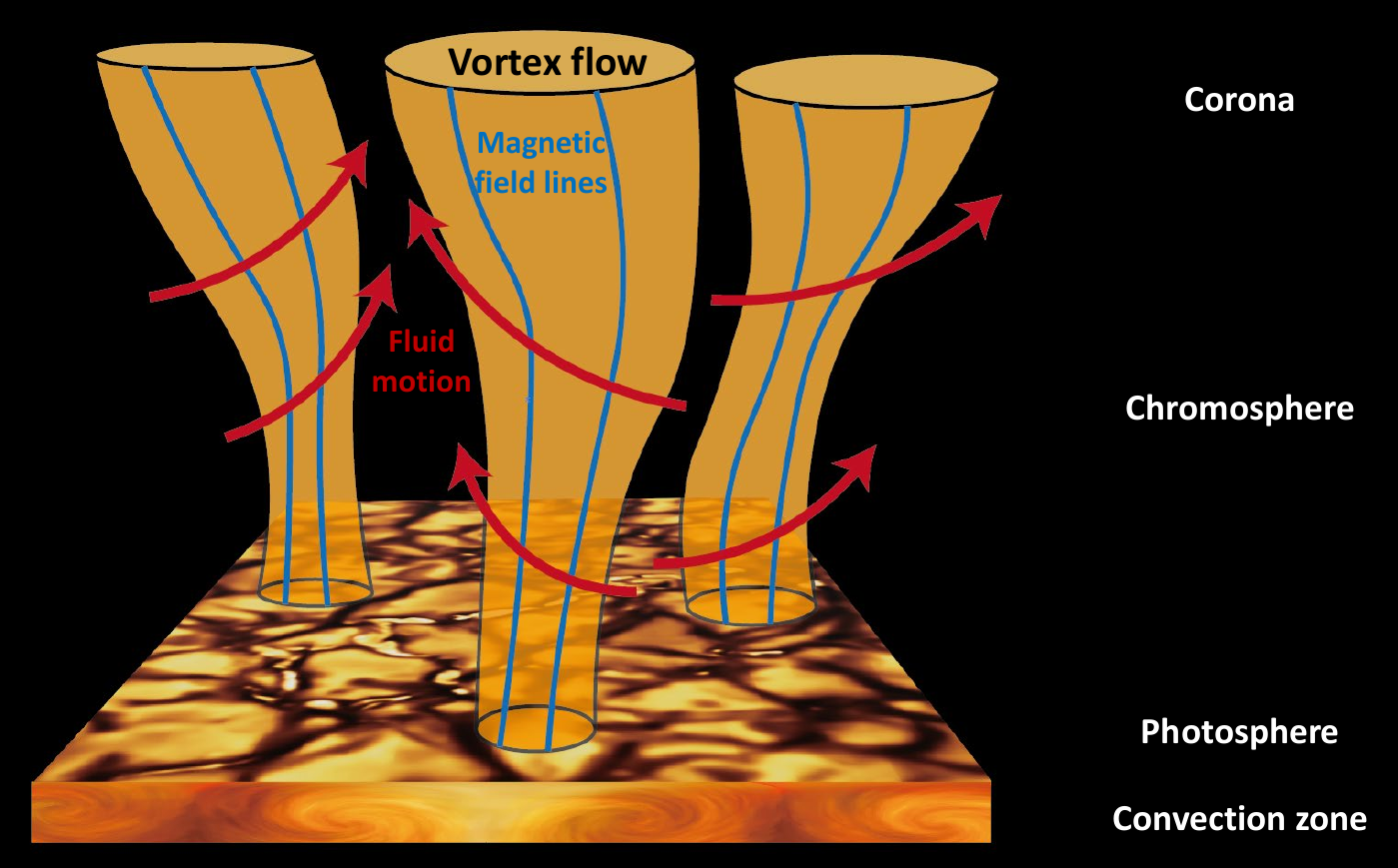}
	\caption{Schematic illustration of vortex flows in the solar atmosphere. Vortices extend from the convection zone through the photosphere and chromosphere into the corona. Blue lines represent magnetic field lines within the vortices, and red arrows indicate the direction of fluid motion.}
	\label{fig:sun}
\end{figure}

Recent high-resolution observations and magnetoconvective simulations of the solar surface and atmosphere have revealed the widespread presence of vortical structures in the solar atmosphere \citep{Silva2020Solar, Breu2023Swirls, Tziotziou2023Vortex}. These vortices are comparable in scale to convective granules and form continuous structures through which a significant portion of the Poynting flux \citep{Silva2024Magnetohydrodynamic} traverses the chromosphere via vortex tubes, establishing magnetic connectivity between the photosphere and the corona, as illustrated schematically in figure~\ref{fig:sun}. These vortices are believed to serve as possible channels for the transport of mass, momentum, and energy from the convection zone to the upper solar atmosphere \citep{Wedemeyer2012Magnetic}, extending from the solar surface up to the lower corona. Their formation and evolution are closely tied to the dynamics of solar convective turbulence and its interactions with intergranular magnetic fields. Vortices with differing rotational directions carry plasma, couple with magnetic fields, extract energy from convective turbulence, and interact in complex ways with large-scale magnetic phenomena in the solar atmosphere, such as sunspots \citep{Bello2012Shear} and coronal mass ejections \citep{Török2013Initiation, Borovsky2006Eddy}.

Vortices and magnetic flux tubes are fundamental coherent structures in MHD systems, playing a central role in the transport of energy and helicity \citep{Moffatt1969The, Moffatt1992Helicity, Chui1995The}. These structures serve as idealized models for investigating the dynamics of complex MHD flows \citep{Titov1999Basic}, with flux tubes being especially prominent in studies of the solar corona \citep{Berger2009Self,Ricca2013New,Pontin2022Magnetic}. From the Sun’s interior to its surface, convection-driven MHD turbulence gives rise to intricate interactions between vortex tubes and magnetic flux tubes. In such environments, the velocity and magnetic fields are strongly coupled \citep{Wedemeyer2012Magnetic}, making the study of their interplay essential for uncovering the physical mechanisms behind MHD structure formation and dynamo action. Models that explicitly couple vortex tubes and magnetic flux tubes provide a valuable framework for exploring these mechanisms.

At high Reynolds numbers, both vorticity and magnetic fields are nearly frozen into the conducting fluid, moving like non-diffusive tracers. Under the analogy between vorticity and magnetic induction, Helmholtz’s theorem and Alfvén's theorem \citep{Davidson2015Turbulence} describe this frozen-in behavior in ideal fluids: the former states that vortex tubes are materially conserved in inviscid flows, preserving their topological structure as they move with the fluid, while the latter asserts that magnetic flux tubes remain embedded in perfectly conducting plasmas.

In real physical systems, however, finite viscosity and magnetic resistivity break these ideal constraints, enabling the reconfiguration of vortical and magnetic structures through reconnection and dissipation. The breakdown of topological invariants governs energy transfer pathways in MHD systems, enabling critical processes such as vortex reconnection \citep{Kida1994Vortex, vanRees2012Vortex, Yao2022Vortex}, vortex merging \citep{Le2022Viscous,Shen2022Topological}, vortex instabilities and cascades \citep{Leweke2016Dynamics,Ostilla2021Cascades,McKeown2020Turbulence}, magnetic reconnection \citep{Li2016Magnetic, Shen2022The, Hao2021Magnetic}, and magnetic splitting \citep{Xiong2020Effects, Kang2025Effects}, all of which redistribute energy across scales. These mechanisms are further modulated by the relative dominance of Lorentz and inertial forces, quantified by the interaction parameter $N_i$ \citep{Davidson2013Turbulence,Kivotides2018Interactions,Kivotides2019Interactions}, which scales the Lorentz force against the inertial force.

Numerical simulations \citep{Nordlund1992Dynamo, Rempel2023Lagrangian, Warnecke2023Numerical} and space-based observations \citep{Tziotziou2023Vortex} of magnetohydrodynamic (MHD) flows have revealed strong couplings between intense vortex tubes and magnetic flux tubes. However, in actual astrophysical plasmas, limited observational capabilities constrain direct analysis of their fluid dynamical behavior. Consequently, model-based studies of vortex–magnetic tube interactions offer a valuable testbed for scientific investigation. \citet{Kivotides2007Magnetic} demonstrated that filamentary vortex structures can sustain dynamo action, leading to the formation of flattened magnetic tubes. Expanding on this, \citet{Kivotides2018Interactions, Kivotides2019Interactions} showed that initially uncoupled vortex tubes and magnetic rings could evolve into magnetic sheets or helical structures under varying interaction parameters. More recently, \citet{Kang2025Effects} explored how interactions between knotted magnetic tubes and vortex tubes affect magnetic splitting, revealing that vortex structures can inhibit the formation of Lorentz-force-induced vortex dipoles in curved magnetic tubes, thereby suppressing magnetic fragmentation. The dynamics observed in these studies are highly sensitive to the initial configurations of the vortex and magnetic structures, which often diverge from those found in realistic astrophysical settings. Moreover, in the Sun and similar turbulent environments, the strengths of co-located vorticity and magnetic fields vary considerably. Despite this, systematic investigations of vortex–magnetic tube couplings analogous to solar vortex flows across a broad range of interaction parameters ($N_i$) remain scarce. 
This gap leaves open critical questions: How do variations in force balances influence instability mechanisms in MHD flows? How do variations in force balances alter the pathways and efficiency of energy transfer in MHD turbulence?
How do variations in force balances shape the emergent large-scale morphology of MHD flow structures?

To advance the understanding of complex MHD phenomena in solar environments, this study investigates vortex–magnetic interactions using co-located antiparallel vortex and magnetic flux tubes modeled after those observed in the solar atmosphere. In such settings, the interplay between vorticity and magnetic fields, both approximately frozen into the plasma, plays a critical role in structuring solar dynamics. This work analyzes how their dynamics are dictated by the competition between vorticity-induced flows and Lorentz forces and explores three distinct regimes: vortex-dominated joint reconnection, instability-triggered cascade, and Lorentz-induced vortex disruption. 
The remainder of this paper is organized as follows. 
\S2 describes the initial configuration and numerical simulation setup.
In \S3, we begin with the evolution of pure vortex or magnetic tubes.
\S4 discusses the three regimes of vortex–magnetic interactions across varying $N_i$. 
In \S5, we examine the effects of the interaction parameter on flux transfer.
Finally, \S6 presents our conclusions.

\section{Simulation overview}

\subsection{Initial configuration of antiparallel flux tubes}

\begin{figure}
	\centering
	\includegraphics[width=\linewidth]{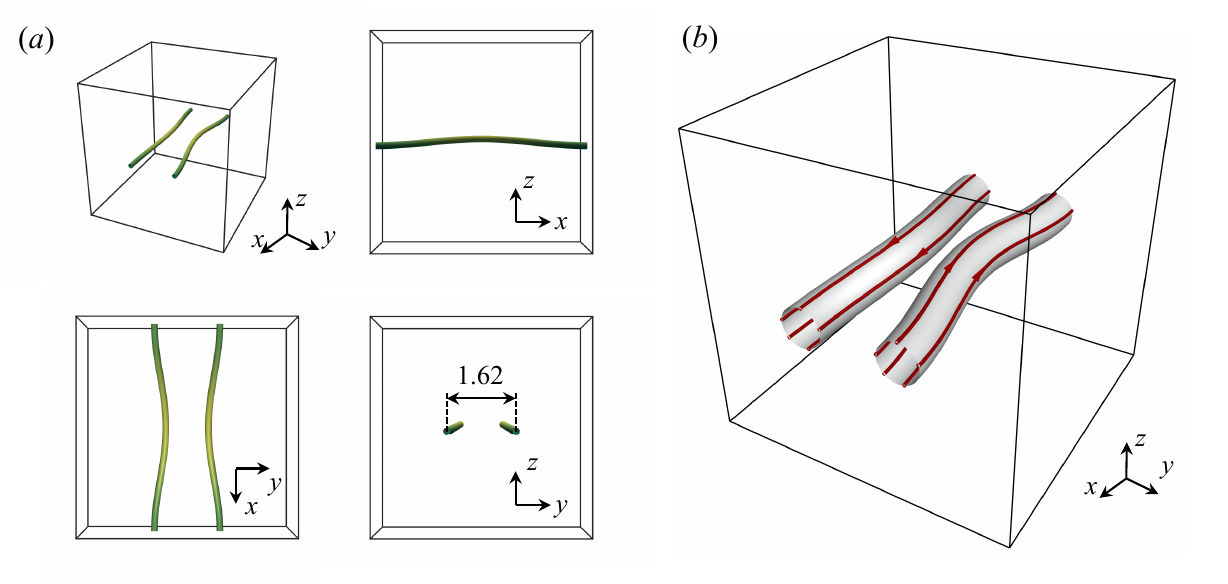}
	\caption{Initial configuration of antiparallel flux tubes. (a) Initial configuration of the centerlines of flux tubes, shown in perspective, side, top, and front views (ordered from left to right, top to bottom). (b) Initial vortex or magnetic surfaces of the antiparallel flux tubes. Some attached field lines (red) are integrated from points on these surfaces.}
	\label{fig:initial}
\end{figure}

The initial configurations of both vortex tubes and magnetic tubes are similar to the standard antiparallel setups commonly used in previous studies \citep{Melander1989Cross, vanRees2012Vortex, Yao2019A} and serve as a suitable model for solar atmospheric vortices, as shown in figure~\ref{fig:sun}. The parametric equations for the centerlines of the two flux tubes are given by
\begin{equation}\label{eq:centerline}
	\boldsymbol{c}(s) = \left( \pm s, y_c(s), z_c(s) \right),
\end{equation}
where
\begin{equation}\label{eq:pery}
	y_c(s) = y_0 + p f(s) \sin \alpha,
\end{equation}
\begin{equation}\label{eq:perz}
	z_c(s) = z_0 + p f(s) \cos \alpha.
\end{equation}
Here, $(y_0, z_0)$ represent the initial positions of the unperturbed vortex centroids, $s$ is the arc-length parameter, $p$ is the amplitude of the axis displacement, and $\alpha$ is the inclination angle.  
For the symmetric pair of antiparallel tubes, we set $(x_0, y_0) = (\mp 0.81, 0)$, $\alpha = \pm \pi/3$, $p = 0.2$, and $f(s) = \cos(s)$. The initial configuration of centerlines is shown in figure~\ref{fig:initial}.

Building upon previously developed geometric construction framework~\citep{Xiong2019Construction,Shen2023Role,Shen2024Designing}, we generate tailored configurations of the vorticity field $\boldsymbol{\omega}$ and the magnetic field $\boldsymbol{b}$ to represent vortex tubes and magnetic flux tubes with given centerline topology, tube thickness, and flux strength. 
We employ a curvilinear cylindrical coordinate system $(s, r, \theta)$, where $s$ denotes the arc length along the tube's centerline $\boldsymbol{c}(s)$, and $(r, \theta)$ are the local polar coordinates in the cross-sectional plane. In this coordinate system, the fields are prescribed as
\begin{equation}\label{eq:tubeconstruction}
	\left\lbrace
	\begin{aligned}
		&\boldsymbol{\omega}(s, r, \theta) = \Gamma_0\, f_t(r)\, \boldsymbol{e}_{s}, \\
		&\boldsymbol{b}(s, r, \theta) = \Gamma_{m0}\, f_t(r)\, \boldsymbol{e}_{s},
	\end{aligned}
	\right.
\end{equation}
where $\boldsymbol{e}_{s} \equiv \mathrm{d}\boldsymbol{c}/\mathrm{d}s$ is the unit tangent vector along the tube centerline, and $\Gamma_0$ and $\Gamma_{m0}$ denote the initial flux of vorticity and magnetic induction, respectively. These quantities define the global strength of the respective fields along the tube.
The radial distribution of both fields within the tube cross-section is modeled by a Gaussian kernel
\begin{equation}\label{eq:Gaussian}
	f_t(r) = \frac{1}{2\pi \sigma_{c0}^2} \exp\left( -\frac{r^2}{2 \sigma_{c0}^2} \right),
\end{equation}
where the standard deviation $\sigma_{c0} = 0.25$ characterizes the initial effective thickness of the flux tubes. This smooth, localized profile ensures well-defined, confined tubular structures, suitable for both analytical and numerical studies.

We ensure that the antiparallel pairs of vortex tubes and magnetic flux tubes are initialized with exactly the same tube thickness and placed at precisely the same initial positions. Under such a configuration, both the vorticity and magnetic field obey their respective frozen-in principles. Small sinusoidal perturbations introduced via \eqref{eq:centerline} lead to distinct response modes in the two fields. In vortex tubes, the deformation triggers Crow instability and reconnection through induced motion \citep{Melander1989Cross}. In contrast, magnetic tubes respond through curvature-induced Lorentz forces, which drive tube splitting \citep{Kang2025Effects}. This setup enables the charged fluid within the flux tubes to exhibit a competitive interplay between inertial forces and Lorentz forces.

For the initial conditions, we prescribe the vorticity flux of each antiparallel vortex tube pair as  
\begin{equation}
	\Gamma = \Gamma_0 = \int_{\mathcal{S}} \boldsymbol{\omega} \cdot \boldsymbol{n} \, \mathrm{d}S = 1,
\end{equation}
where $\boldsymbol{\omega}$ is the vorticity vector and $\mathcal{S}$ denotes a cross-sectional surface of the tube. Subsequently, we superimpose magnetic flux tubes with various magnetic flux strengths given by  
\begin{equation}
	\Gamma_m = \Gamma_{m0} = \int_{\mathcal{S}} \boldsymbol{b} \cdot \boldsymbol{n} \, \mathrm{d}S,
\end{equation} 
at the location of the vortex tubes, where $\Gamma_{m0}$ ranges from 0.02 to 1. Here, $\boldsymbol{b}$ is the dimensionless magnetic induction field, normalized by the Alfvén velocity $\boldsymbol{b} = \boldsymbol{B}/\sqrt{\mu \rho}$, in which $\boldsymbol{B}$ is the magnetic field, $\rho$ is the fluid density, and $\mu$ is the magnetic permeability of free space.

In this study, we focus primarily on flows characterized by a vortex Reynolds number  
\begin{equation}
	\Rey = \frac{u \ell}{\nu} \sim \frac{\Gamma_0}{\nu} = 2000,
\end{equation}
and a magnetic Prandtl number $Pr_m = \nu/\eta = 1$, so that the magnetic Reynolds number becomes  
\begin{equation}
	R_m = \frac{u \ell}{\eta} \sim \frac{\Gamma_0}{\eta} = \Rey.
\end{equation}
Here, $u$ and $l$ denote characteristic velocity and characteristic length, respectively.
This ensures that vorticity diffusion due to kinematic viscosity $\nu$ is balanced with magnetic diffusion, where the magnetic diffusivity is given by $\eta = (\sigma \mu)^{-1}$ and $\sigma$ is the electrical conductivity.

The interaction parameter~\citep{Davidson2015Turbulence,Kivotides2018Interactions} quantifies how strongly the magnetic field influences the inertial dynamics of flow structures. It is defined as
\begin{equation}
	N_i = \frac{\nabla \times (\text{Lorentz force})}{\nabla \times (\text{Inertial force})} = \frac{\sigma B^2 \ell}{\rho u},
\end{equation}
and represents the dimensionless ratio measuring the relative magnitude of the Lorentz force to the inertial force. A higher value of $N_i$ indicates a stronger magnetic modulation of the vortex-dominated turbulence, marking a departure from purely hydrodynamic behavior.
Substituting $B = b \sqrt{\mu \rho}$ and choosing the tube thickness $\sigma_c$ as the characteristic length scale, we obtain  
\begin{equation}\label{eq:Ni}
	N_i = \frac{(b \ell^2)^2}{(u \ell)(\sigma \mu)^{-1} \ell^2} \sim \frac{\Gamma_m^2}{\Gamma \eta \sigma_c^2} .
\end{equation}
In the current study, our choice of parameters spans a broad range of interaction strengths, corresponding to $N_i$ values from $12.8$ to $32000$, thus covering regimes from vorticity-dominated dynamics to those dominated by Lorentz forces.

In the numerical implementation, the construction in \eqref{eq:tubeconstruction} is realized within a periodic domain of size $(2\pi)^3$. This is achieved through a differential geometry-based mapping technique~\citep{Xiong2019Construction,Xiong2020Effects,Shen2023Role,Shen2024Designing}, which transforms coordinates from a curvilinear cylindrical system $(s, r, \theta)$ to a Cartesian grid. This approach facilitates the generation of strictly solenoidal vorticity or magnetic fields, thereby ensuring that the initial conditions used in DNS are physically consistent and well-posed. This numerical construction has been demonstrated to be highly robust in generating smooth vorticity fields and magnetic fields even in cases of turbulence and complex flux tube interactions~\citep{Shen2024Designing}. The initial configuration of the constructed flux tubes is visualized in figure~\ref{fig:initial}(b) via vortex surfaces and magnetic flux surfaces. Each vortex line or magnetic field line lies precisely embedded within the corresponding vortex or magnetic surface.

The velocity field $\boldsymbol{u}$ is then obtained from the prescribed vorticity field $\boldsymbol{\omega}$ via the Biot–Savart (BS) law, which is implemented in Fourier space for computational efficiency. The relationship is given by
\begin{equation}
	\boldsymbol{u} = \mathcal{F}^{-1}\left(\frac{i \boldsymbol{k} \times \hat{\boldsymbol{\omega}}}{|\boldsymbol{k}|^2}\right),
\end{equation}
where $\mathcal{F}^{-1}$ denotes the inverse Fourier transform, $\boldsymbol{k}$ is the wavevector in Fourier space, and $\hat{\boldsymbol{\omega}} = \mathcal{F}(\boldsymbol{\omega})$ represents the Fourier transform of the vorticity field.

\subsection{Direct numerical simulation of MHD flows}

We solve the three-dimensional incompressible MHD equations \citep{gurnett2017introduction} with a constant unit density $\rho$. The first is the fluid momentum equation
\begin{equation}\label{eq:u}
	\frac{\partial \boldsymbol{u}}{\partial t} + \boldsymbol{u} \cdot \nabla \boldsymbol{u} = -\nabla \left(\frac{p}{\rho}\right) + \boldsymbol{j} \times \boldsymbol{b} + \nu \nabla^2 \boldsymbol{u}.
\end{equation}
The evolution of the magnetic field $\boldsymbol{b}$ is governed by the induction equation
\begin{equation}\label{eq:b}
	\frac{\partial \boldsymbol{b}}{\partial t} = \nabla \times (\boldsymbol{u} \times \boldsymbol{b}) + \eta \nabla^2 \boldsymbol{b}.
\end{equation}
The incompressibility conditions $\nabla \cdot \boldsymbol{u} = 0$ and $\nabla \cdot \boldsymbol{b} = 0$ are enforced. Here, $\boldsymbol{x} = (x, y, z)$ denotes the Cartesian spatial coordinates, $\boldsymbol{j} = \nabla \times \boldsymbol{b}$ is the current density, $p$ is the pressure.
Furthermore, the evolution of vorticity field $\boldsymbol{\omega} = \nabla \times \boldsymbol{u}$ can be described by
\begin{equation}\label{eq:omega}
	\frac{\partial \boldsymbol{\omega}}{\partial t} = \nabla \times (\boldsymbol{u} \times \boldsymbol{\omega}) + \nu \nabla^2 \boldsymbol{\omega} + \nabla \times (\boldsymbol{j} \times \boldsymbol{b}).
\end{equation}
Notably, aside from the Lorentz force contribution in the final term, the structure of this equation mirrors that of the magnetic induction equation~\eqref{eq:b}, highlighting the formal analogy between vorticity dynamics and magnetic field evolution in MHD.

Direct numerical simulations of MHD flows are performed in a triply periodic cubic domain $\Omega$, discretized using $N^3$ uniform grid points. 
We employ a pseudo-spectral method to solve Eqs.~\eqref{eq:u} and \eqref{eq:b} in the symmetric Elsässer form $\boldsymbol{z}^{\pm} = \boldsymbol{u} \pm \boldsymbol{b}$ \citep{Elsasser1950The}. Aliasing errors are mitigated using the two-thirds truncation rule, restricting the maximum resolved wavenumber to $k_{\max} \approx N/3$. Time advancement of Fourier coefficients is achieved using a second-order Runge-Kutta scheme. 
The time step is chosen to ensure the Courant–Friedrichs–Lewy (CFL) numbers below 0.5 for both the velocity and magnetic fields, thereby ensuring numerical stability and accuracy throughout the simulation.
This numerical solver has been used and validated in magnetic reconnection~\citep{Hao2021Magnetic}.
Time is non-dimensionalized as $t^*=t /\left(2 \pi l_s^2 / \Gamma_0\right)$, where $l_s = 1.62$ is the initial separation between vortex tubes, and $\Gamma_0$ is the initial circulation of either tube. This scaling implies that $t^*$ approximately represents the time taken for the undisturbed vortex dipole to travel a distance $l_s$.

To ensure that the grid resolution can fully resolve the flow evolution, the grid number $N$ is carefully selected in the range of 512 to 1024 based on the case under consideration. A grid convergence analysis is detailed in Appendix~\ref{app:grid} for a vortex–magnetic joint reconnection case at different Reynolds numbers which has the most demanding resolution requirements due to the strong gradients in both velocity and magnetic fields.

\section{Evolution of pure antiparallel vortex or magnetic tubes}\label{sec:purevortexmagnetic}

We begin by examining two limiting cases of the interaction parameter: one involving pure antiparallel vortex tubes without a magnetic field, and the other involving pure antiparallel magnetic flux tubes embedded in an initially stationary, electrically conducting fluid. These correspond to $N_i = 0$ and $N_i \rightarrow \infty$, respectively. In both cases, we apply the same small perturbation~\eqref{eq:centerline} to the tubes, arranged in the same spatial configuration. The two reference problems are designed to separately isolate the intrinsic dynamics of the perturbed vortex tubes and magnetic tubes. For the pure vortex tube case, the circulation is set to $\Gamma = 1$, and the kinematic viscosity is set to $\nu = 5 \times 10^{-4}$, giving a vortex Reynolds number $\Rey = 2000$. For consistency, the pure magnetic tube case is initialized with a magnetic flux of $\Gamma_m = 1$ and a magnetic diffusivity of $\eta = 5 \times 10^{-4}$.

\begin{figure}
	\centering
	\includegraphics[width=\linewidth]{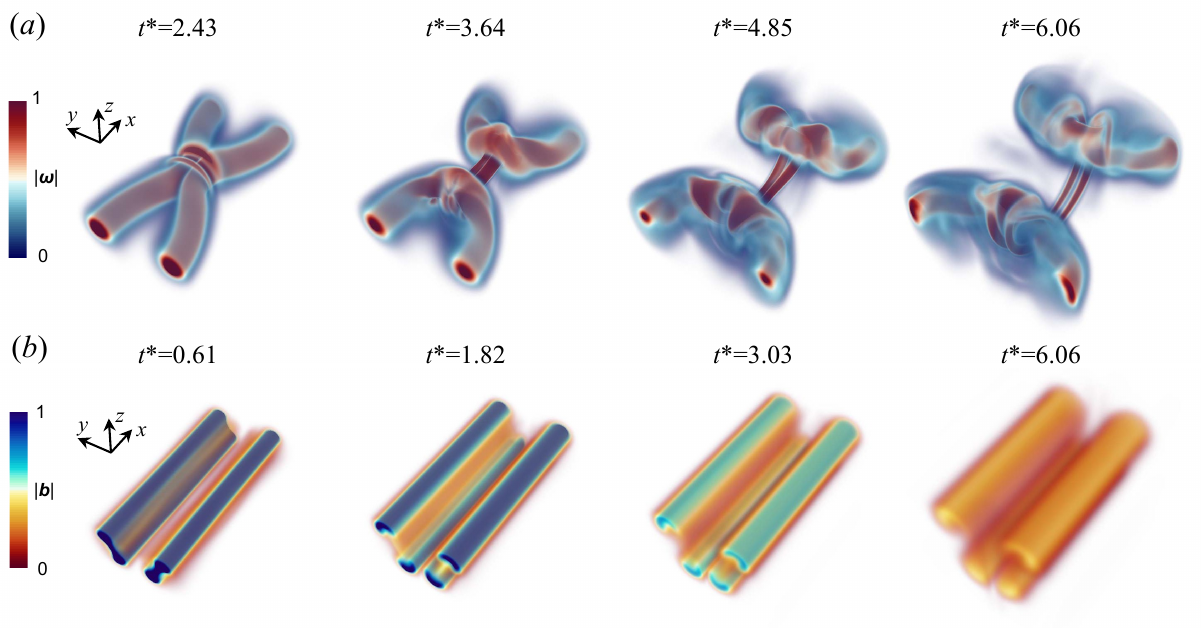}
	\caption{Evolution of flow structures for (a) pure vortex reconnection at $N_i=12.8$ and (b) pure magnetic splitting at $N_i \rightarrow \infty$, visualized through volume rendering of (a) vorticity magnitude $|\boldsymbol{\omega}|$ and (b) magnetic induction magnitude $|\boldsymbol{b}|$.}
	\label{fig:pure}
\end{figure}

\begin{figure}
	\centering
	\includegraphics[width=0.9\linewidth]{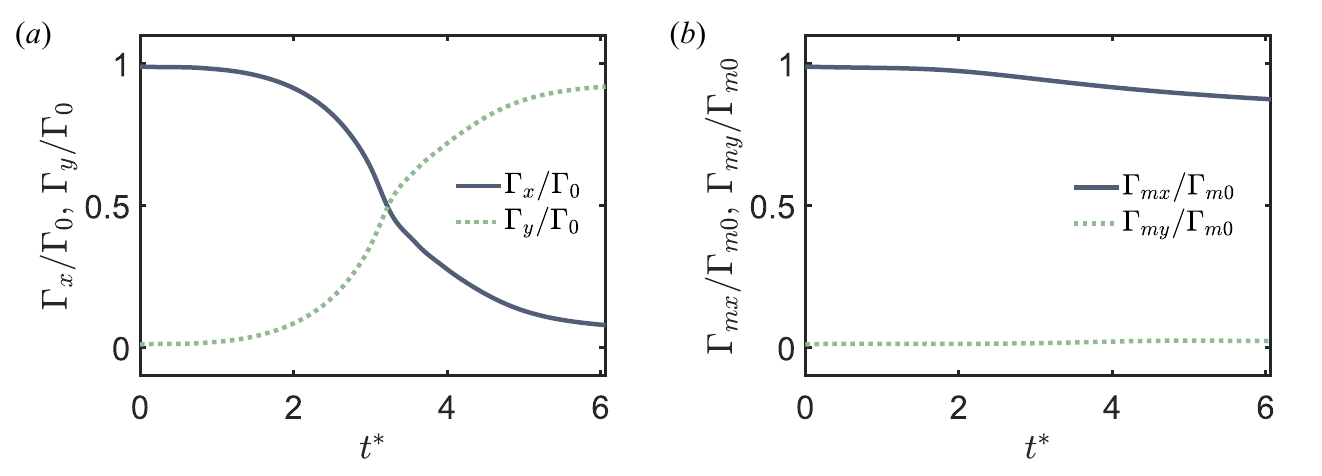}
	\caption{Flux transfer in (a) pure vortex reconnection and (b) magnetic splitting. Shown is the time evolution of (a) vorticity flux and (b) magnetic flux through half of the $x = 0$ (solid lines) and $y = 0$ (dashed lines) planes, normalized by the initial flux in the $x = 0$ plane.}
	\label{fig:pureflux}
\end{figure}

In the evolution of antiparallel vortex tubes, the absence of magnetic fields reduces the MHD equations to pure hydrodynamics, and standard vortex reconnection is thus triggered, as extensively studied in previous works \citep{Melander1989Cross, vanRees2012Vortex}. Due to mutual induction, the antiparallel vortex tubes rise together and approach each other, eventually undergoing topological change through reconnection (see figure~\ref{fig:pure}(a)). Thread-like structures remain after reconnection, and the flux tubes subsequently separate. The vorticity flux, initially aligned with the $x$-direction, remains nearly conserved but is transferred to the transverse $y$-direction (figure~\ref{fig:pureflux}(a)).

In the pure magnetic tube configuration, we consider magnetic flux tubes embedded in an initially stationary fluid. At the initial moment, the system contains only a magnetic field distribution, with no externally imposed flow field. Although the induction equation \eqref{eq:b} does not contain an explicit Lorentz force term, the magnetic field configuration is not force-free. The imbalance in the initial magnetic field generates a self-consistent Lorentz force $\boldsymbol{F}_{\text{L}} \equiv \boldsymbol{j} \times \boldsymbol{b}$, which in turn drives motion according to the momentum equation \eqref{eq:u}.
Subsequently, the magnetic field evolves via the field amplification and diffusion terms implicitly contained in the induction equation. Due to the absence of vortex-induced motion, the flux tubes exhibit neither significant net motion in the $z$-direction nor convergence in the $y$-direction. As shown in figure~\ref{fig:pure}(b), the system rapidly transitions to a quasi-two-dimensional state. Each magnetic tube undergoes axial splitting into two thinner tubes, driven by small initial perturbations and the action of the Lorentz force~\citep{Xiong2020Effects, Kang2025Effects}.

This Lorentz-force-driven splitting originates from the perturbed curvature $\kappa$ of the initial magnetic field lines. The nonzero curl of the Lorentz force~\citep{Kang2025Effects},
\begin{equation}\label{eq:source}
	\boldsymbol{\nabla} \times \boldsymbol{F}_L = \boldsymbol{\nabla} \times \left(b^2 \kappa \boldsymbol{N}\right),
\end{equation}
acts as a source term in the vorticity equation \eqref{eq:omega}, where $(\boldsymbol{T}, \boldsymbol{N}, \boldsymbol{B})$ denotes the Frenet–Serret frame of the magnetic field lines with $\boldsymbol{T}$, $\boldsymbol{N}$, and $\boldsymbol{B}$ representing the local tangent, normal, and binormal directions, respectively. For planar two-dimensional perturbations \eqref{eq:centerline}, equation \eqref{eq:source} simplifies to $\boldsymbol{\nabla} \times \boldsymbol{F}_L = -(\kappa \boldsymbol{B} \cdot \boldsymbol{\nabla} b^2) \boldsymbol{T}$, which forms vortex dipoles in the cross-section of the flux tubes and drives splitting along the binormal direction $\boldsymbol{B}$ of the field lines.

However, in the absence of mutual advection, the antiparallel flux tubes do not approach each other, and magnetic reconnection is inhibited. As a result, the system enters a long-term dissipative phase dominated by magnetic diffusion.
Consistent with Alfvén’s theorem, the magnetic flux initially aligned with the $x$-direction is well preserved during the early stage, when the isolated magnetic tube undergoes splitting without contacting the antiparallel tube (figure~\ref{fig:pureflux}(b)). In the later stage, magnetic diffusion causes the tube radius to increase, gradually bringing the oppositely directed lower flux tubes into contact and leading to slow flux cancellation.

These two benchmark cases demonstrate the vortex and magnetic "frozen-in" behavior of the fluid, respectively. This contrast highlights the fundamental differences between fluid motion driven by magnetic tubes and by vortex tubes: the collision and merging of vortex tubes versus the stretching and splitting of magnetic tubes. It also foreshadows the competing or antagonistic interplay that arises when both structures are present and equally perturbed. Specifically, the former tends to facilitate the approach and reconnection of antiparallel structures, while the latter appears to inhibit such processes. This antagonism is influenced by the relative strength of the magnetic and vortex tubes, that is, by the value of the interaction parameter $N_i$.

\section{Regimes of vortex–magnetic interactions}\label{sec:regimes}

The magnitude of the interaction parameter determines the relative dominance of vortex dynamics versus Lorentz-force-driven evolution in the charged fluid system. Shifts in this balance can drive the system toward fundamentally different dynamical behaviors, playing a crucial role in shaping flow evolution and governing energy transfer pathways. In the present study, we focus on numerical simulations spanning a wide range of interaction parameters, $N_i = 12.8 \sim 32000$, corresponding to a fixed initial vortex flux $\Gamma_{0} = 1$ and magnetic flux values of the flux tubes in the range $\Gamma_{m0} = 0.02 \sim 1$. Specifically, the interplay between vortical motions and magnetic fields gives rise to three distinct regimes of vortex–magnetic interactions: vortex-dominated joint reconnection, instability-triggered cascade, and Lorentz-induced vortex disruption. These regimes reflect how vortices and magnetic fields modulate the fluid dynamics depending on the dominance of inertia or electromagnetic forces, leading to markedly different topological and energetic outcomes.

\subsection{Vortex-dominated joint reconnection for low $N_i$}\label{subsec:reconnection}

\begin{figure}
	\centering
	\includegraphics[width=\linewidth]{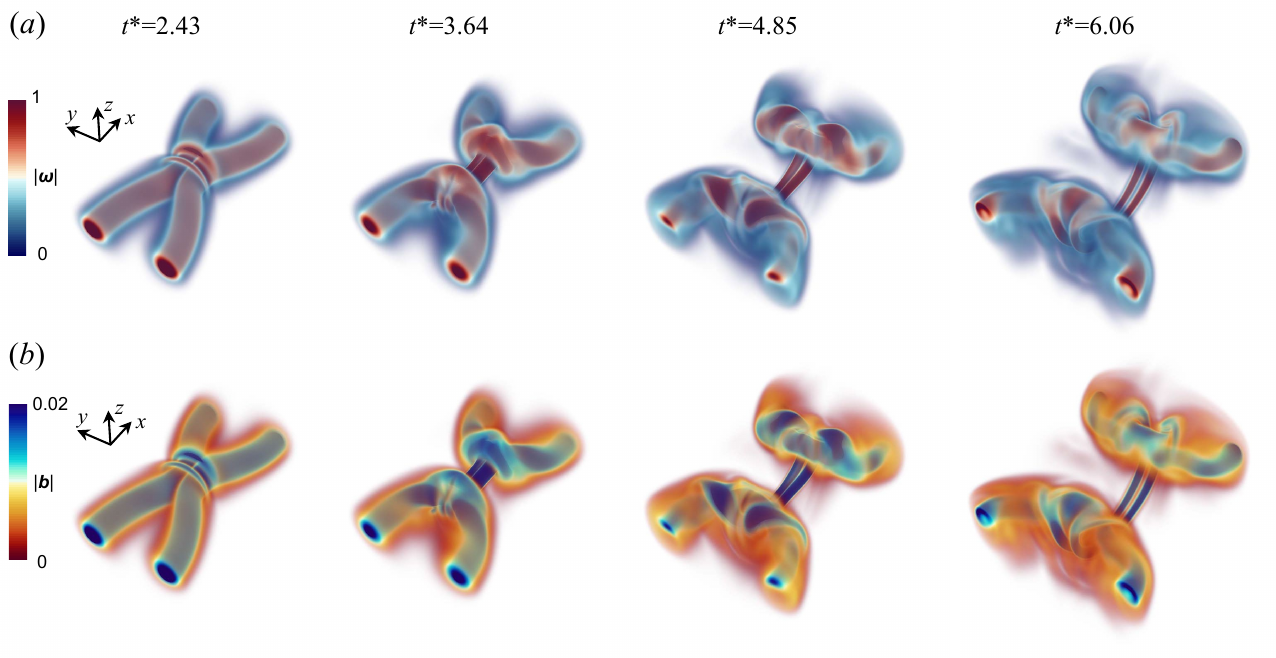}
	\caption{Evolution of flow structures for vortex–magnetic joint reconnection at $\Rey=R_m=2000$ and $N_i=12.8$, visualized through volume rendering of (a) vorticity magnitude $|\boldsymbol{\omega}|$ and (b) magnetic induction magnitude $|\boldsymbol{b}|$.}
	\label{fig:rec}
\end{figure}

In the regime of low interaction parameter $N_i$, vortex dynamics prevail. Small initial perturbations can trigger the Crow instability in vortex tubes, which serves as the primary mechanism driving reconnection at low $N_i$. Perturbations to magnetic flux tubes can lead to their straightening or splitting. However, when $N_i$ is small, the Lorentz force is weak relative to the vortex-induced flow and plays little role in initiating either the instability or reconnection. Under the influence of the frozen-in law, magnetic field lines are advected by the motion of vortex structures. Figure~\ref{fig:rec}(a,b) illustrates the evolution of both vortex and magnetic structures during a vortex–magnetic joint reconnection at $\Rey=R_m=2000$ and $N_i=12.8$. The vorticity and magnetic field are tightly coupled, exhibiting a strong Lagrangian coherence.

Vortex dynamics, governed by the frozen-in nature of vorticity, dominate the flow. Vortex tubes entrain magnetic flux tubes, lifting them upward and initiating simultaneous vortex and magnetic reconnections. This joint reconnection process imparts to magnetic reconnection the same complex structural features typically seen in vortex reconnection—namely, bridging and threading~\citep{Melander1989Cross,Yao2022Vortex}, where the main flux tubes reconnect and leave behind fine-scale thread-like structures. This contrasts with typical magnetic reconnection, which is generally much simpler, often fast and clean~\citep{Hao2021Magnetic}. Classical models, often based on current sheets \citep{Yamada2010Magnetic,Pontin2022Magnetic}, capture large-scale reconnection but neglect vorticity and interaction effects, thus missing finer structural features.

\begin{figure}
	\centering
	\includegraphics[width=0.95\linewidth]{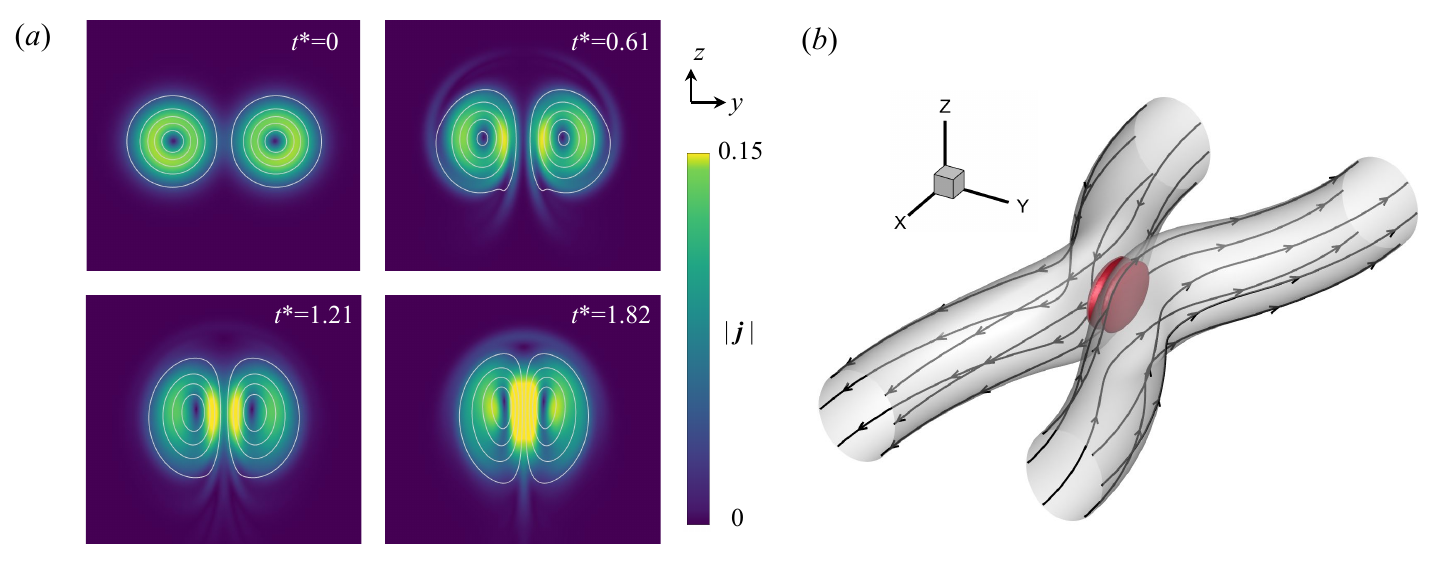}
	\caption{Formation of current sheets during reconnection.
		(a) Evolution of current density magnitude $|\boldsymbol{j}|$ shown as colour contours, with vorticity magnitude $\boldsymbol{\omega}$ overlaid as isolines on the $y$–$z$ symmetry plane before and during the reconnection process.
		(b) Three-dimensional structure of the current sheet and magnetic flux tubes at $t^* = 1.82$. The white isosurface represents the magnetic field magnitude $|\boldsymbol{b}| = 0.01$, outlining the outer boundary of the flux tubes. Selected magnetic field lines are shown on this surface. The red isosurface corresponds to $|\boldsymbol{j}| = 0.15$, highlighting the high-current region seen in panel (a).}
	\label{fig:current}
\end{figure}

Magnetic reconnection occurs primarily within current sheets, where electric current density concentrates. As reconnection proceeds, the current increases sharply at the reconnection site. Figure~\ref{fig:current}(a) shows the evolution of the current density magnitude $|\boldsymbol{j}|$ on the $y$–$z$ symmetry plane. Vorticity isolines mark the positions and boundaries of the flux tubes. Driven by vortex advection, high current density layers emerge at the contact interface of the flux tubes. According to Ampère’s law, this corresponds to intense magnetic field variation $\nabla \times \boldsymbol{b} =  \boldsymbol{j}$. Figure~\ref{fig:current}(b) presents the three-dimensional structure and spatial location of the current sheets. A disk-shaped current sheet forms at the interface between approaching vortex tubes. Electric current penetrates this sheet from below and exits above, while magnetic field lines reconnect within the sheet.

\begin{figure}
	\centering
	\includegraphics[width=0.95\linewidth]{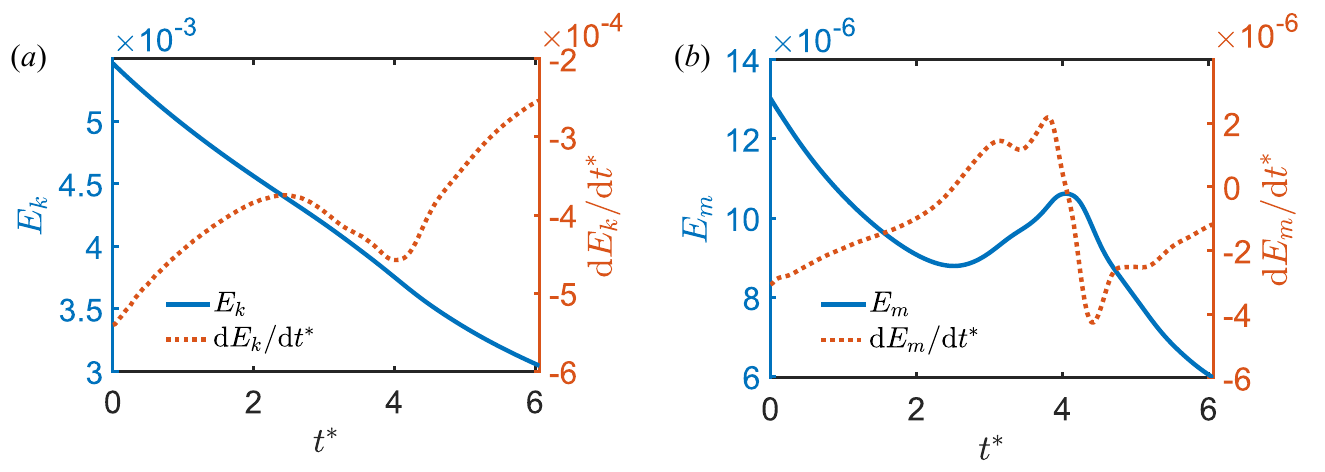}
	\caption{Temporal evolution of volume-averaged energies for the reconnection case with $N_i = 12.8$. (a) Kinetic energy and its time derivative.	(b) Magnetic energy and its time derivative.}
	\label{fig:recE}
\end{figure}

The evolution of the flow also governs the energy balance of the system. Under a low interaction parameter $N_i$, the kinetic energy significantly exceeds the magnetic energy. As shown in figure~\ref{fig:recE}(a), the kinetic energy decreases monotonically throughout the global flow, and the influence of the magnetic field on this decay is negligible. Vortex reconnection is a highly dissipative process, rapidly reducing the kinetic energy.
Surprisingly, the magnetic energy increases during the reconnection process and reaches its maximum after reconnection is completed (see figure~\ref{fig:recE}(b)). The total electromechanical energy in the system, defined as the sum of kinetic and magnetic energies, decreases due to Ohmic heating and viscous dissipation
\begin{equation}
	\frac{\mathrm{d} E}{\mathrm{d} t} = \frac{\mathrm{d}}{\mathrm{d} t} \int_V \left[ \frac{\boldsymbol{u}^2}{2} + \frac{\boldsymbol{b}^2}{2} \right] dV = - \int_V \left( \eta\boldsymbol{j}^2 \right) dV - 2 \nu \int_V S_{ij} S_{ij} , dV,
\end{equation}
where $S_{ij}$ is the strain-rate tensor. Given that kinetic and magnetic energy are the only forms of energy contributing to this balance, the observed increase in magnetic energy implies a conversion of kinetic energy into magnetic energy via vortex-induced magnetic reconnection. This phenomenon can be viewed as a manifestation of the dynamo effect, which is often of central interest in the interiors of the Sun and stars. This contrasts sharply with the canonical magnetic reconnection observed at the solar surface, where oppositely directed magnetic field lines break and reconnect, rapidly converting magnetic energy into kinetic energy. On the Sun, this released energy drives high-energy particle jets and generates radiation.

\begin{figure}
	\centering
	\includegraphics[width=0.9\linewidth]{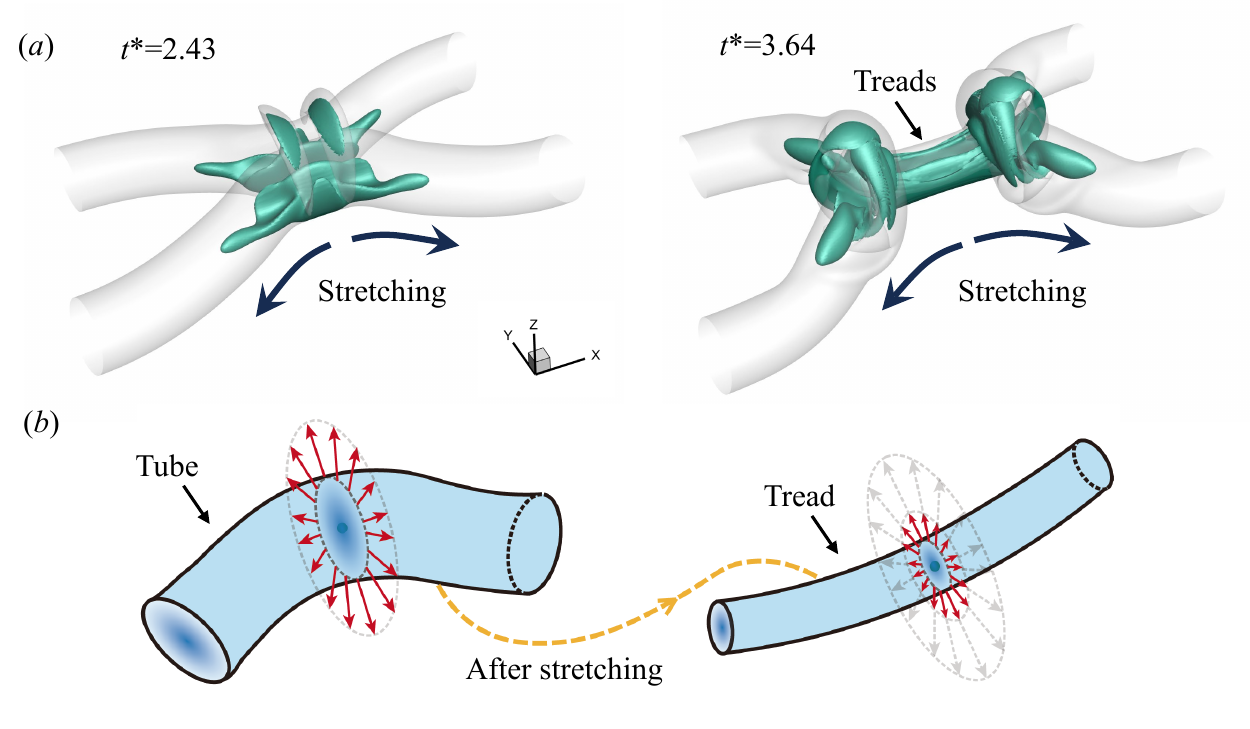}
	\caption{Regions and mechanisms of magnetic energy growth during vortex-magnetic joint reconnection.
		(a) The green isosurface of $\mathcal{W}_{\text{L}} = -1 \times 10^{-4}$ encloses regions where the Lorentz force performs negative work, located near the reconnection sites and stretched as reconnection proceeds.
		(b) Schematic of a magnetic flux tube and the Lorentz force on its magnetic surface before and after stretching. Red arrows indicate the Lorentz force at the cross-section. As flux tubes are stretched into slender threads, the Lorentz force performs negative work, contributing to magnetic energy growth.}
	\label{fig:recEt}
\end{figure}

Understanding the physical mechanism behind the vortex-induced growth of magnetic energy is crucial for gaining deeper insight into the dynamo effect. The evolution of magnetic energy is governed by the following equation
\begin{equation}
	\frac{\mathrm{d}}{\mathrm{d} t} \int_V \left( \frac{\boldsymbol{b}^2}{2} \right) dV = - \int_V \left[ (\boldsymbol{j} \times \boldsymbol{b}) \cdot \boldsymbol{u} \right] dV - \int_V \left( \eta\boldsymbol{j}^2 \right) dV,
\end{equation}
where the term involving the Lorentz force represents the work done on the plasma. The conversion from kinetic to magnetic energy occurs when the Lorentz force performs negative work $\mathcal{W}_{\text{L}} = \int_{V}[(\boldsymbol{j} \times \boldsymbol{b}) \cdot \boldsymbol{u}] \mathrm{d} V <0$. Figure~\ref{fig:recEt}(a) highlights the regions with negative $\mathcal{W}_{\text{L}}$ during reconnection, localizing the magnetic energy growth to flux tubes and slender thread-like structures near the reconnection region.

For a Gaussian-distributed magnetic flux tube \eqref{eq:tubeconstruction}, the Lorentz force is given by
\begin{equation}
	\boldsymbol{F}_{\text{L}} = (\nabla \times \boldsymbol{b}) \times \boldsymbol{b} = \frac{\Gamma_m^2 r}{4 \pi^2 \sigma_c^6} \exp \left( -\frac{r^2}{\sigma_c^2} \right) \boldsymbol{e}_r,
\end{equation}
where the Lorentz force points radially outward within the flux tube.
During reconnection, vortex tubes stretch the magnetic tubes into slender threads, and as the material surfaces retract, the Lorentz force continuously performs negative work. This sustained mechanism facilitates the conversion of kinetic energy into magnetic energy. Figure~\ref{fig:recEt}(b) schematically illustrates this physical process.

\begin{figure}
	\centering
	\includegraphics[width=\linewidth]{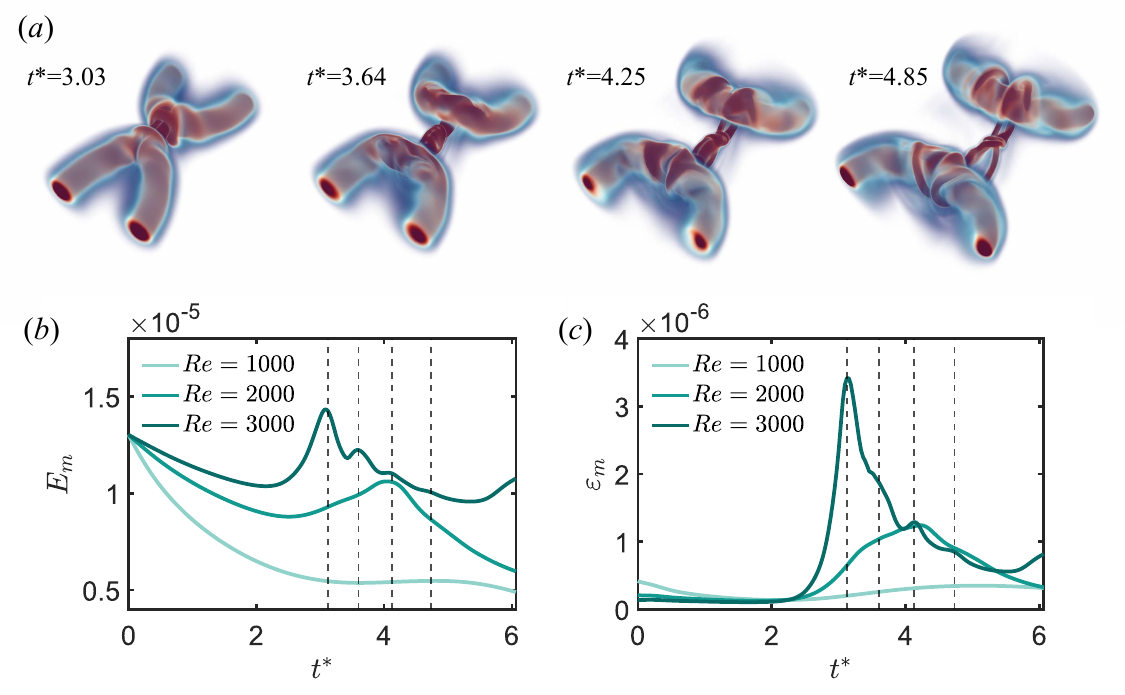}
	\caption{Reynolds number effects in the vortex–magnetic joint reconnection.
		(a) Evolution of flow structures at $\Rey = R_m = 3000$, visualized by volume rendering of the vorticity magnitude $|\boldsymbol{\omega}|$.
		(b) Temporal evolution of magnetic energy for different Reynolds numbers.
		(c) Temporal evolution of magnetic energy dissipation rate for different Reynolds numbers.
		The dashed lines in (b) and (c) indicate the time of peak values of the corresponding quantities, occurring at $t^* = 3.12$, 3.6, 4.12, and 4.73 from left to right.}
	\label{fig:recRe}
\end{figure}

The Reynolds number effect on reconnection plays a critical role in modulating the efficiency and characteristics of the vortex-induced dynamo mechanism. Much like in standard vortex reconnection~\citep{Yao2019A}, increasing the Reynolds number leads to multistage reconnection of fine-scale threads. As illustrated in figure~\ref{fig:recRe}(a), during reconnection at a higher Reynolds number ($\Rey = 3000$), secondary reconnection occurs among the residual threads following the initial reconnection of the primary flux tubes. This process generates progressively finer vortex and magnetic structures. Notably, as the Reynolds number increases, symmetry breaking emerges at late times. This results from mutual induction and stretching between the residual threads, which form a dipole structure at their upper ends \citep{Yao2019A}. The adjacent, elongated tails create a planar, jet-like shear layer that becomes highly susceptible to Kelvin–Helmholtz instability, a mechanism also observed in previous high-Reynolds-number hydrodynamic reconnection studies (e.g., \citet{vanRees2012Vortex,Beardsell2016Investigation}) and now shown to arise in the MHD context as well.

At the same time, the peak value of magnetic energy increases with the Reynolds number, as shown in figure~\ref{fig:recRe}(b). This indicates that the dynamo effect becomes more pronounced at higher Reynolds numbers, with enhanced conversion of kinetic energy into magnetic energy. The occurrence of successive reconnection events at small scales is further evidenced by multiple peaks in the temporal evolution of both magnetic energy and magnetic dissipation rate (see figure~\ref{fig:recRe}(c)).

\begin{figure}
	\centering
	\includegraphics[width=\linewidth]{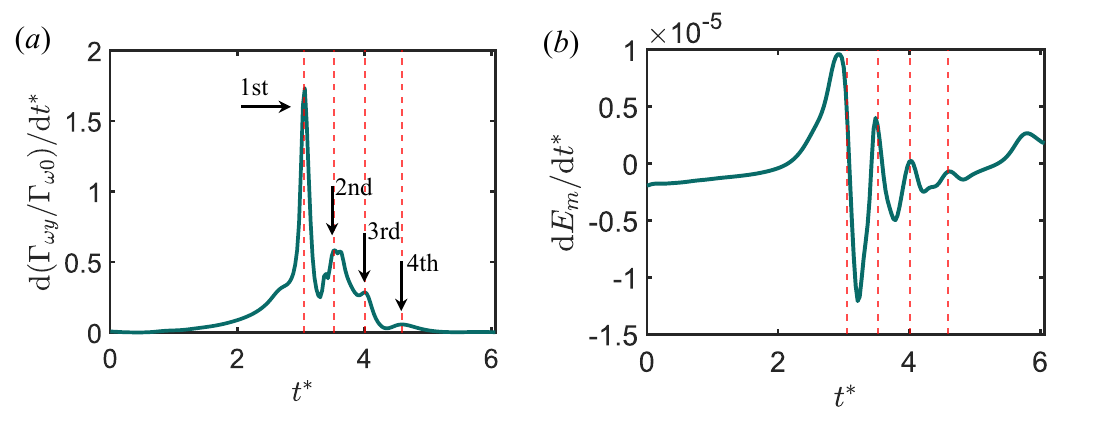}
	\caption{Temporal derivatives of (a) $y$-direction flux transfer and (b) magnetic energy for vortex–magnetic joint reconnection at $\Rey = R_m = 3000$ and $N_i = 19.2$. The red dashed lines in panels (a) and (b) indicate the characteristic times of successive reconnection events, specifically at $t^* = 3.05,\ 3.52,\ 4.01,\ 4.59$.}
	\label{fig:recRederi}
\end{figure}

To establish the correlation between the multi-peak behaviour of magnetic energy and the occurrence of secondary reconnections, we analyzed the temporal derivatives of the $y$-direction flux transfer and the magnetic energy, as shown in figure~\ref{fig:recRederi}. The time derivative of the $y$-flux exhibits multiple peaks, indicating that flux transfer occurs in a discontinuous manner through successive reconnection events. These peak times mark the characteristic moments of each reconnection stage and are indicated by red dashed lines in figure~\ref{fig:recRederi}. When compared with the derivative of the magnetic energy, we observe that the reconnection times align closely with the peaks in magnetic energy growth rate. This strong temporal correspondence suggests a causal relationship: each reconnection event directly contributes to a new phase of magnetic energy growth. These secondary reconnections are fundamentally similar to the primary event, only occurring at smaller scales. Therefore, the mechanism of magnetic energy amplification, arising from flux tube stretching, applies equally to each secondary reconnection. As the Reynolds number continues to increase, it is foreseeable that higher-order generations of reconnection will develop, ultimately forming a turbulent cloud avalanche composed of a tangled network of fine-scale vortex and magnetic structures. This results in sustained magnetic energy growth and a cascade of magnetic energy down to ever-smaller spatial scales.

\subsection{Instability-triggered cascade for moderate $N_i$}\label{subsec:cascade}

\begin{figure}
	\centering
	\includegraphics[width=\linewidth]{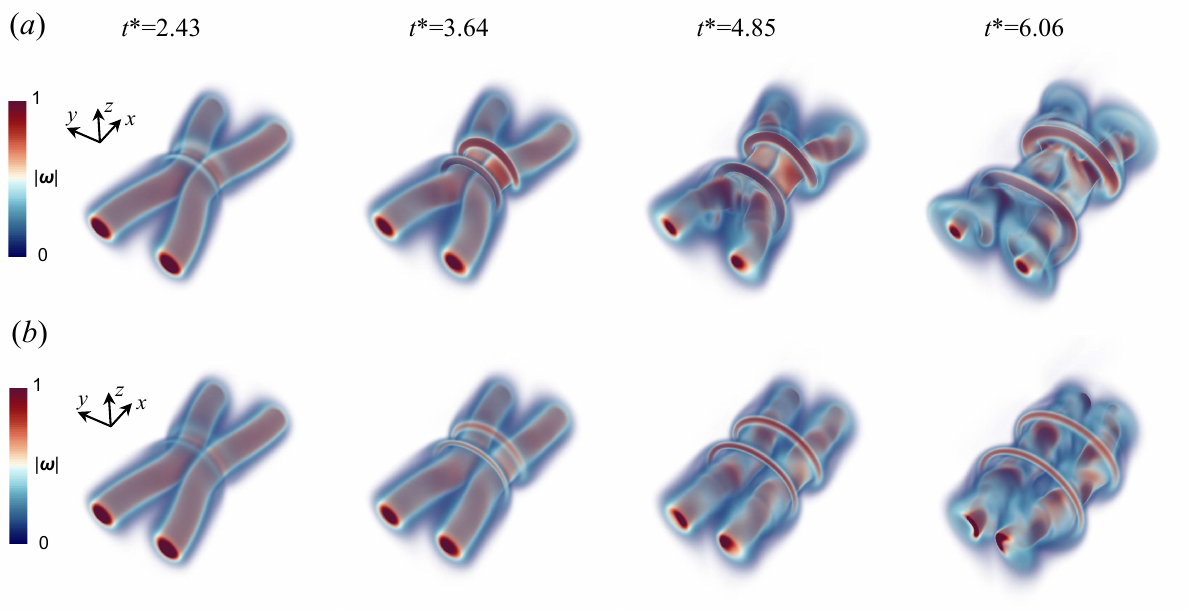}
	\caption{Evolution of flow structures for instability-triggered cascade at (a) $N_i=115.2$ and (b) 204.8, visualized through volume rendering of vorticity magnitude $|\boldsymbol{\omega}|$.}
	\label{fig:cascade}
\end{figure}

A moderate interaction parameter implies a dynamic balance between vortex dynamics and magnetodynamics. Figure~\ref{fig:cascade} illustrates the evolution of flow structures at moderate values of $N_i = 115.2$ and $204.8$. The magnetic and vortex structures remain visually indistinguishable (not shown), suggesting continued coupling between vorticity and magnetic field lines.

Unlike the complete reconnection observed at low interaction parameters, flux tubes at intermediate interaction parameters initiate reconnection upon contact but separate again after transferring only a portion of their flux. This behavior indicates that flux tubes undergo only partial reconnection in this regime.
The portion of the tube that successfully reconnects decreases as the interaction parameter increases. The reconnected segments of the flux tubes reorient perpendicularly relative to their original alignment, forming secondary filamentary structures. These secondary antiparallel filaments, both vortex and magnetic, arise from localized reconnection at the contact region, where oppositely directed fields cancel and partial fluid exchange occurs. Meanwhile, as the unreconnected portions gradually separate from the contact region, the filaments are stretched and slightly displaced. This leads to a morphological transition of the flux tubes from an initial X-shape to an O-shaped configuration (see figure~\ref{fig:cascade}(b)), signifying a dynamic balance between magnetic tension and vorticity-induced attraction. The charged fluid in this regime oscillates between the attractive forces generated by vortex dynamics and the damping effect of the magnetic field. Kelvin waves~\citep{Leweke2016Dynamics} and Alfvén waves \citep{cramer2011physics} play a crucial role in this process, inducing instabilities that cause the flux tubes to oscillate and deform as if plucked like a stretched string.

\begin{figure}
	\centering
	\includegraphics[width=0.9\linewidth]{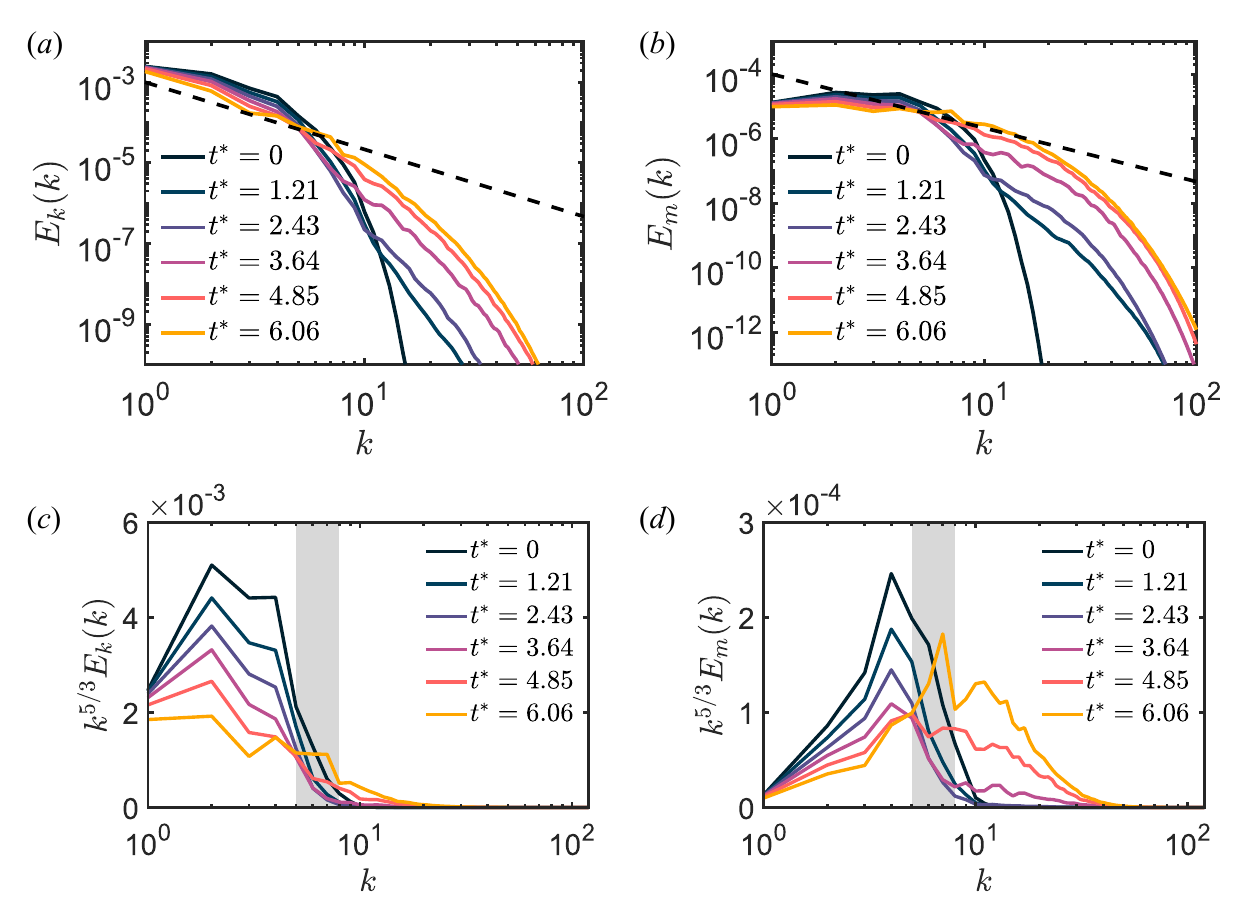}
	\caption{Temporal evolution of (a) kinetic energy spectra and (b) magnetic energy spectra at $N_i = 115.2$ and $\Rey=2000$ with the $k^{-5/3}$ slope shown as dashed lines. Panels (c) and (d) show the corresponding compensated spectra for kinetic and magnetic energy, respectively, highlighting the overlapping scaling regions ($k\approx 5\text{–}8$).}
	\label{fig:cascadeEsp}
\end{figure}

These wave-induced instabilities, combined with nonlinear interactions, give rise to secondary antiparallel filaments. This mechanism iterates, initiating an energy cascade. Analogous to the cascades triggered by Crow instability and elliptical instability in classical vortex dynamics~\citep{McKeown2020Turbulence}, the flow develops progressively finer structures over time. Both the kinetic and magnetic energy spectra gradually approach the characteristic turbulent scaling, following a Kolmogorov-like spectrum of $k^{-5/3}$, consistent with observations of solar wind turbulence~\citep{Goldstein1995Magnetohydrodynamic,Podesta2007Spectral}, as illustrated in figure~\ref{fig:cascadeEsp}(a,b). 

Although the kinetic and magnetic energy spectra both exhibit regions consistent with a $k^{-5/3}$ scaling, there is a slight offset between them (see the compensated spectra in figure~\ref{fig:cascadeEsp}(c,d)). This mismatch likely reflects differences in the cascade processes of kinetic and magnetic energy, arising from the interplay between vortex distortion and Alfvén wave interactions \citep{Schekochihin2022MHD,Muller2005Spectral}. Nevertheless, the scaling regimes of two spectra partially overlap around $k\approx 5 \text{-} 8$. Notably, our system has no external energy input, yet the spectra clearly show an increase in small-scale energy over time. This energy growth can only result from a cascade from larger scales, indicating that energy is transferred to smaller scales through repeated generations of filamentation driven by instabilities. As a result, the initially coherent vortex tubes continuously fragment into finer filaments, enriching the small-scale structure of the flow.

\subsection{Lorentz-induced vortex disruption for high $N_i$}\label{subsec:damping}

\begin{figure}
	\centering
	\includegraphics[width=\linewidth]{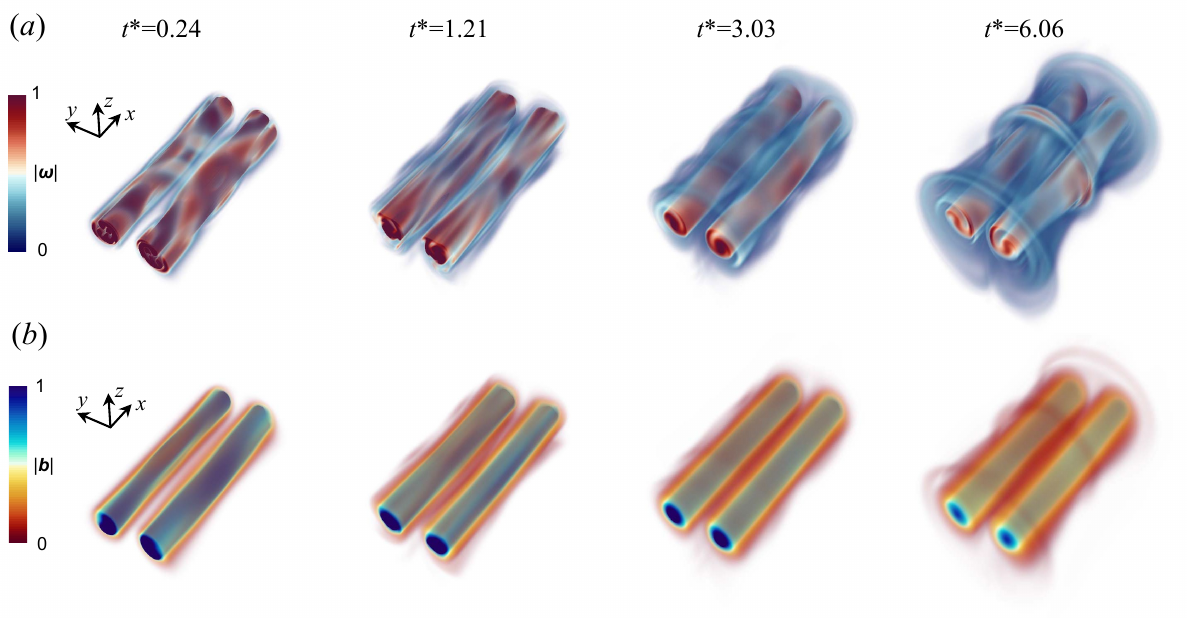}
	\caption{Evolution of flow structures for Lorentz-induced vortex disruption at $N_i=32000$, visualized through volume rendering of (a) vorticity magnitude $|\boldsymbol{\omega}|$ and (b) magnetic induction magnitude $|\boldsymbol{b}|$.}
	\label{fig:damping}
\end{figure}

When the magnetic field becomes dominant, the system enters the regime of large interaction parameters. Figure~\ref{fig:damping} illustrates the evolution of flow structures, both vortex and magnetic, at a high interaction parameter of $N_i = 32000$. In this regime, the initially coupled vortex and magnetic structures begin to decouple, exhibiting clear distinctions in their behavior. 
For the vortex structures, the Lorentz force rapidly disrupts their coherence, tearing apart the vortex cores and deforming them into loosely wound spiral shapes. Moreover, the Lorentz force counteracts the induced motions that would otherwise draw vortex tubes together. As a result, the tubes become isolated, preventing the mutual interactions between antiparallel pairs, such as reconnection and energy cascade, as previously discussed.
In contrast, during the early stages, the flux tubes behave similarly to the pure magnetic tubes described in Section~\ref{sec:purevortexmagnetic}. They rapidly straighten due to magnetic tension. However, the condensation effect of the surrounding vortex field suppresses their tendency to split or fragment. Consequently, over longer timescales, the magnetic tubes largely retain their structure, undergoing gradual dissipation with minimal morphological change.

\begin{figure}
	\centering
	\includegraphics[width=0.9\linewidth]{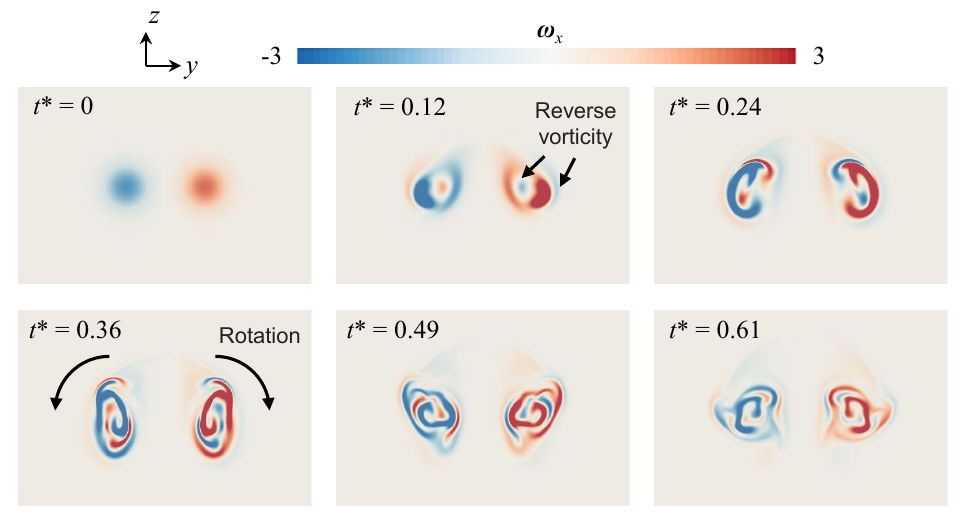}
	\caption{Evolution of axial vorticity $\boldsymbol{\omega}_x$ contours on the $y$–$z$ symmetry plane before and during the vortex disruption process at $N_i=32000$.}
	\label{fig:core}
\end{figure}

Figure~\ref{fig:core} presents the contours of axial vorticity, $\boldsymbol{\omega}_x$, on cross-sections of the flux tubes, providing a clearer view of the vortex disruption process. Initially, the vortex tubes are stretched and distorted by the Lorentz force, which induces shear and consequently generates reverse vorticity. This reverse vorticity then interleaves with the original vorticity inside the flux tubes, undergoing a global rotation along the streamlines and ultimately becoming nested and rolled into a complex, spiral structure.

\begin{figure}
	\centering
	\includegraphics[width=0.85\linewidth]{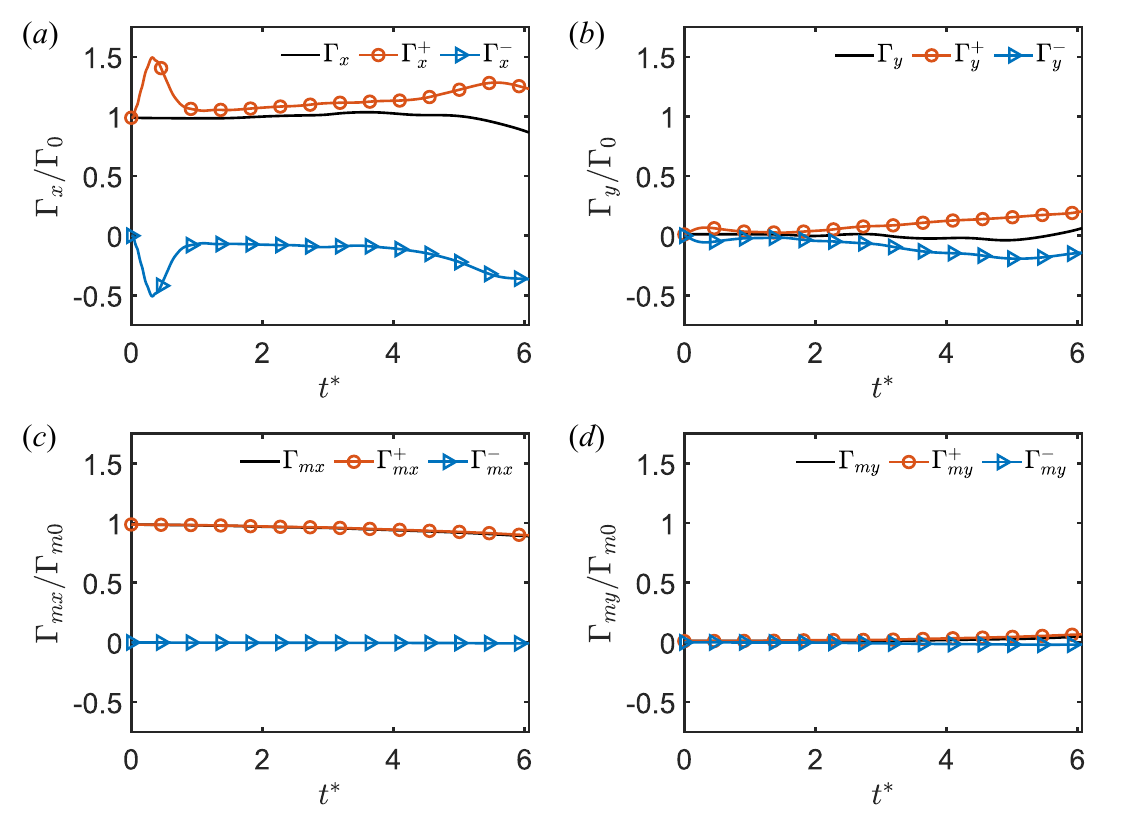}
	\caption{Temporal evolution of (a,b) vorticity flux and (c,d) magnetic flux for $N_i=32000$ case, each decomposed into positive and negative contributions, through half of the (a,c) $x = 0$ and (b,d) $y = 0$ planes. All fluxes are normalized by the initial flux in the $x = 0$ plane.}
	\label{fig:decom}
\end{figure}

Despite the disruption of vortex tubes, the circulation of both the vortex and magnetic fluxes remains conserved under Helmholtz's law. Figure~\ref{fig:decom} illustrates the decomposition of vortex and magnetic flux into their positive and negative components. The reverse axial vorticity generated by shearing appears rapidly during vortex core disruption and is subsequently dissipated (figure~\ref{fig:decom}(a)). Simultaneously, the positive circulation grows and dissipates in tandem. However, no significant generation of reverse magnetic field is observed during this process. Furthermore, the negligible flux in the $y$-direction confirms that interactions between antiparallel tubes are suppressed. As vortex core dissipation progresses, any mutual interactions of antiparallel tubes evolve only slowly.

Although the vorticity and magnetic field evolution equations in MHD share structural similarities, they are not identical. The vorticity equation~\eqref{eq:omega} includes an additional source term, the curl of the Lorentz force $\nabla \times (\boldsymbol{j} \times \boldsymbol{b})$, which acts as a vorticity source induced by magnetic effects. As the interaction parameter increases (i.e., stronger magnetic field), this term becomes more prominent, allowing the Lorentz force to perturb vortex cores and generate significant reverse vorticity. In contrast, the magnetic induction equation~\eqref{eq:b} lacks a corresponding vorticity-driven source term, so the magnetic field evolution is only weakly affected by the vorticity field. This fundamental asymmetry explains the absence of significant reverse magnetic field generation.

\begin{figure}
	\centering
	\includegraphics[width=\linewidth]{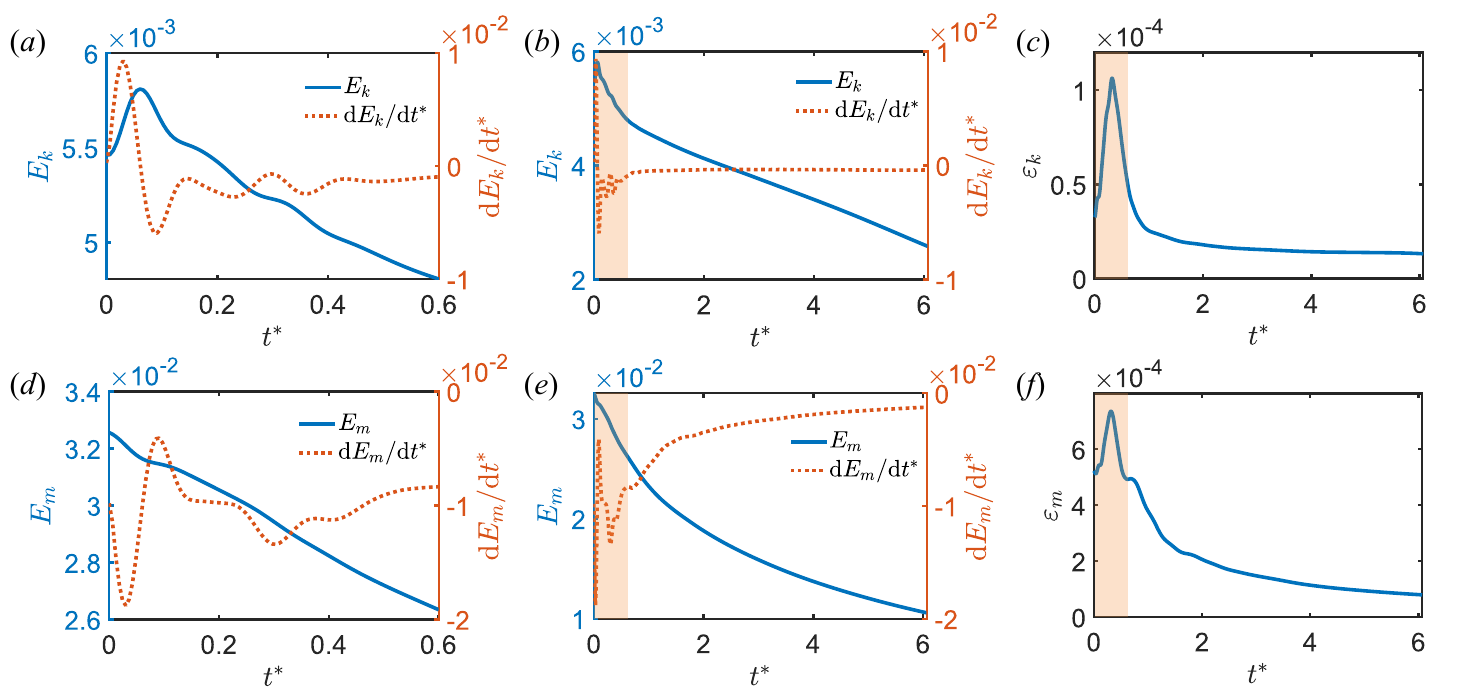}
	\caption{Temporal evolution of volume-averaged quantities for the vortex disruption case with $N_i = 32000$.
		(a) Kinetic energy and its time derivative during the short-term vortex disruption phase.
		(b) Kinetic energy and its time derivative during the long-term dissipation phase.
		(c) Kinetic energy dissipation rate.
		(d) Magnetic energy and its time derivative during the short-term vortex disruption phase.
		(e) Magnetic energy and its time derivative during the long-term dissipation phase.
		(f) Magnetic energy dissipation rate.
		The orange areas in (b-c) and (e-f) highlight the vortex disruption phase.}
	\label{fig:dampingE}
\end{figure}

In the regime of high interaction parameters, the energy transfer mechanisms differ fundamentally from those observed at lower $N_i$. During the initial vortex disruption phase, which occurs over a very short timescale, the Lorentz-force-driven perturbations lead to a partial conversion of magnetic energy into kinetic energy. This is reflected in the rise of kinetic energy and the corresponding decline in magnetic energy (see Figure~\ref{fig:dampingE}(a,d)). Subsequently, strong nonlinear interactions emerge, acting as highly dissipative mechanisms for both vorticity and magnetic field structures (Figure~\ref{fig:dampingE}(c,f)). During this phase, kinetic and magnetic energies undergo oscillatory decay, primarily due to the rotational wrapping and twisting of core structures. After this intense transient period of nonlinear dynamics, both vortex and magnetic tubes settle into a phase of long-term, stable dissipation, as illustrated in Figure~\ref{fig:dampingE}(b,e). This indicates that the flow is constrained and relaxed by the strong magnetic field, with nonlinear interactions and the onset of turbulence effectively suppressed.

\section{Effects of the interaction parameter on flux transfer}\label{sec:effects}

The transfer of flux within flux tubes, specifically the redistribution of vorticity and magnetic field from the initial $x$-direction into transverse directions, serves as a key diagnostic for tracking the topological evolution of the flow. This flux transfer quantifies the dynamic interplay between two competing frozen-in fields: vorticity and magnetic induction. Their competition governs local flow behaviour and underlies the emergence of distinct evolutionary regimes.

\begin{figure}
	\centering
	\includegraphics[width=0.97\linewidth]{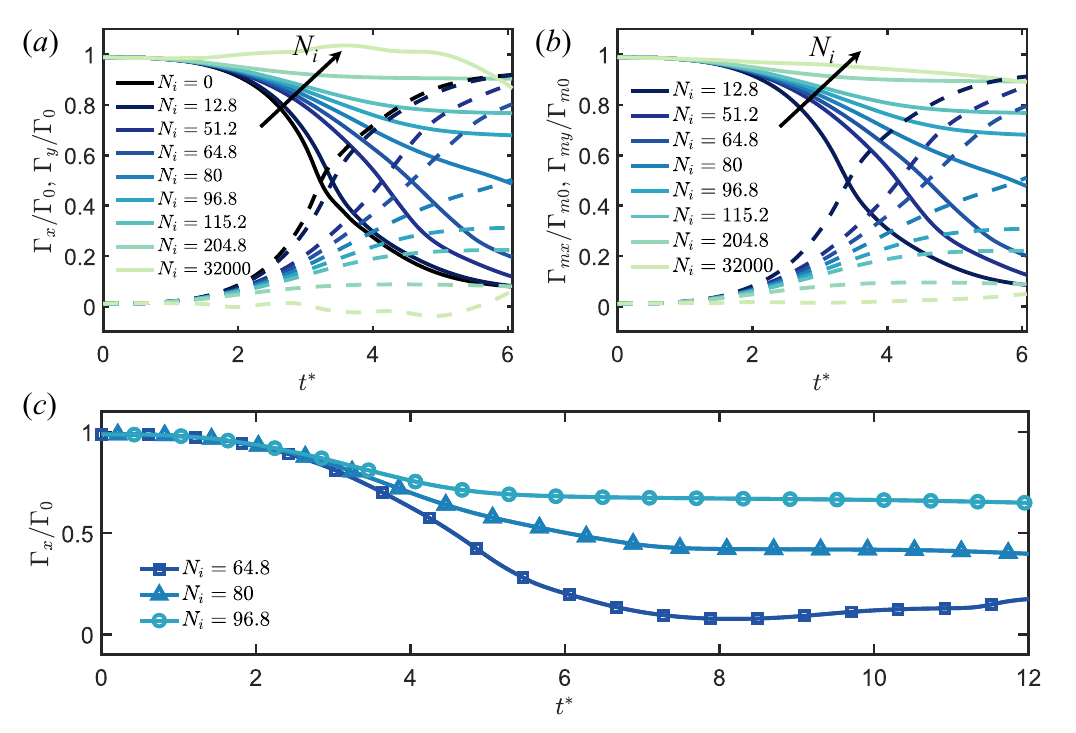}
	\caption{Flux transfer under different interaction parameters at $\Rey=2000$. (a) Time evolution of the vorticity flux through half of the $x = 0$ (solid lines) and $y = 0$ (dashed lines) planes, normalized by the circulation $\Gamma_0$ in the $x = 0$ plane at $t^* = 0$. The color gradient from dark to light represents increasing initial interaction parameters $N_i = 0,\ 12.8,\ 51.2,\ 64.8,\ 80,\ 96.8,\ 115.2,\ 204.8 ,\ 8000,\ 32000$, corresponding to initial magnetic flux values of the flux tubes $\Gamma_{m0} = 0,\ 0.02,\ 0.04,\ 0.045,\ 0.05,\ 0.055,\ 0.06,\ 0.08 ,\ 0.5,\ 1$, respectively. (b) Time evolution of the magnetic flux through half of the $x = 0$ (solid lines) and $y = 0$ (dashed lines) planes, normalized by the initial magnetic flux $\Gamma_{m0}$ in the $x = 0$ plane. (c) Long-time evolution of vorticity flux transfer for three moderate interaction parameters, $N_i = 64.8,\ 80,\ 96.8$.}
	\label{fig:flux}
\end{figure}

Figures~\ref{fig:flux}(a) and (b) respectively present the vortex flux transfer and magnetic flux transfer from the $x$-direction to the $y$-direction under different interaction parameters $N_i=0 \sim 32000$, normalized by their initial values. Except for the extreme case with a very large interaction parameter ($N_i = 32000$), the flux transfers of vorticity and magnetic field exhibit remarkable agreement, as expected from the frozen-in law of vorticity and magnetic fields. This strong correlation indicates that both quantities remain quasi-Lagrangian attached to the fluid elements. The coordinated evolution in the $x$- and $y$-directions further confirms that the sum of the fluxes in the two half-planes ($\Gamma_x + \Gamma_y$ and $\Gamma_{mx} + \Gamma_{my}$) remains conserved, consistent with Helmholtz’s theorem and Alfvén’s theorem, thereby validating the numerical accuracy.

Flux transfer is consistent with the visualization results in section~\ref{sec:regimes}.
For small interaction parameters (e.g., $N_i = 12.8,\ 51.2$), both vorticity and magnetic fluxes are efficiently transferred from the $x$-direction to the $y$-direction within a finite time, indicating that both vortex tubes and magnetic tubes undergo reconnection. As the interaction parameter increases, the magnetic field significantly suppresses the transfer of both vorticity and magnetic flux (see figures~\ref{fig:flux}(a) and (b)), leading to a reduced flux transfer rate and thus inhibiting the reconnection of vortex and magnetic tubes. When $N_i$ becomes large (e.g., $N_i > 200$), $\Gamma_x$ and $\Gamma_{mx}$ remain nearly unchanged, indicating that the flux tubes no longer undergo reconnection. Strong Lorentz forces instead dominate the flow, and the induced motion that would otherwise bring the antiparallel vortex tubes together becomes negligible. At ultra-high $N_i=32000$, we observe partial decoupling between vorticity and magnetic fluxes, indicating the presence of strong nonlinear effects.

Notably, at moderate interaction parameters—where inertial and Lorentz forces are comparable in magnitude—only partial reconnection of the flux tubes occurs within the effective reconnection time window. This behavior stands in sharp contrast to the low interaction parameter regime, where reconnection is fully triggered and the flux is almost entirely transferred. While increasing the interaction parameter generally suppresses the reconnection rate, the transition from full to partial reconnection is not simply a matter of slower dynamics. To confirm that the behavior observed at moderate interaction parameters corresponds to genuinely partial reconnection, rather than a delayed but complete reconnection, we performed long-time evolution studies for three representative cases with intermediate values: $N_i = 64.8, 80, 96.8$. The long-time evolution of vorticity flux transfer is shown in figure~\ref{fig:flux}(c). In all three cases, the flux tubes do not undergo full reconnection at a later time, as might be expected. Instead, somewhat unexpectedly, after an initial partial transfer, both the vorticity and magnetic fluxes plateau, indicating a definitive cessation of the reconnection process.

\begin{figure}
	\centering
	\includegraphics[width=0.99\linewidth]{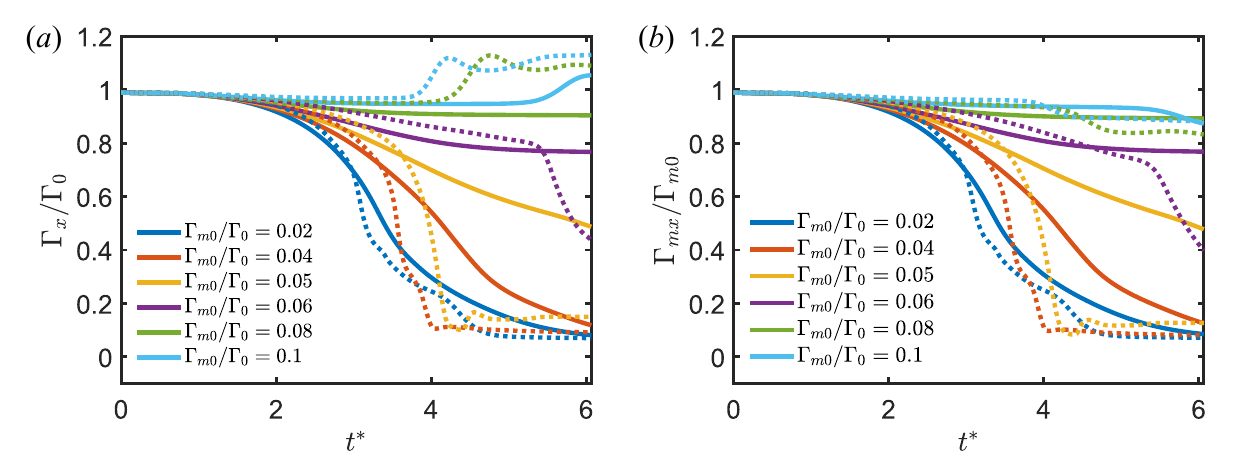}
	\caption{Time evolution of the normalized (a) vortex flux and (b) magnetic flux through half of the $x = 0$ planes at $\Rey=2000$ (solid lines) and $\Rey=3000$ (dotted lines). The same color denotes the same initial condition.}
\label{fig:fluxRe}
\end{figure}

Since the extent and rate of flux transfer may also depend on viscous effects, we further examined how varying the Reynolds number influences the reconnection dynamics. We compared the flux transfer at $\Rey=2000$ and $\Rey=3000$ under identical initial conditions, as shown in figure~\ref{fig:fluxRe}. At lower Reynolds numbers, flux transfer is initially faster due to stronger viscous diffusion, which allows the thicker flux tubes to come into contact earlier. In contrast, during the later stages of evolution, the higher-Reynolds-number cases exhibit faster flux transfer owing to their finer tube thickness. Moreover, as the Reynolds number increases, enhanced nonlinear interactions and the emergence of smaller-scale structures make the flux transfer more intermittent.

In MHD, the steady-state configuration of fluxes plays a critical role in determining structure strength in complex flow fields, such as energy transport channels from the solar convection zone to the upper solar atmosphere \citep{Wedemeyer2012Magnetic}, as shown in figure~\ref{fig:sun}. Here, we attempt to provide a theoretical explanation for the observed partial reconnection and stabilized residual flux under moderate interaction parameters.

During the reconnection process, as oppositely directed vectors of the flux tubes come into contact, the field-line reconnection proceeds from the outer regions inward. This implies that both vorticity and magnetic field within the flux tubes are gradually stripped away from the periphery toward the core. This process is accompanied by a reduction in the characteristic length scale, the tube thickness $\sigma_c$, and a corresponding gradual transfer of flux from the original antiparallel vortex tubes
\begin{equation}\label{eq:Gamma}
	\Gamma = \lambda \Gamma_0
\end{equation}
\begin{equation}\label{eq:Gammam}
	\Gamma_m = \lambda \Gamma_{m0}
\end{equation}
where $\lambda \in (0,1)$ denotes the residual fraction of $x$-directed flux and decreases continuously as reconnection progresses.

The evolution of these parameters directly affects the instantaneous value of the interaction parameter
\begin{equation}\label{eq:Ni2}
	N_i = \frac{\lambda \Gamma_{m0}^2}{\Gamma_0 \eta \sigma_c^2} = \frac{\lambda \sigma_{c0}^2}{\sigma_c^2} N_{i0}
\end{equation}
where $N_{i0}$ is the initial value of the interaction parameter prior to reconnection. This dynamic evolution may ultimately suppress the further development of reconnection.

For a flux tube with a Gaussian distribution \eqref{eq:Gaussian}, the vorticity flux enclosed within a tubular region of radius $r$ from the vortex core is given by
\begin{equation}\label{eq:fluxr}
	\Phi_\omega(r) = \int_0^r 2\pi r \boldsymbol{\omega}(r) \mathrm{d}r  = \Gamma_0 \left[1 - \exp\left(-\frac{r^2}{2\sigma_{c0}^2}\right) \right].
\end{equation}
In the limit of very high Reynolds number, we assume that the initial Gaussian distribution of the flux tube remains frozen during the reconnection process, while the vorticity and magnetic field are progressively eroded from the outer layers inward. We define the effective tube thickness $\sigma_c$ such that the tubular region within $r = \sigma_c$ always contains a fixed fraction $\varepsilon$ of the total residual flux
\begin{equation}\label{eq:tubethickness}
	\Phi_\omega(\sigma_c) = \varepsilon \Gamma,
\end{equation}
where $\varepsilon = 1 - \exp\left(-\frac{1}{2}\right)$, ensuring consistency with the initial Gaussian distribution $\sigma_c(t^* = 0) = \sigma_{c0}$.
Substituting \eqref{eq:Gamma} and \eqref{eq:fluxr} into \eqref{eq:tubethickness}, we obtain
\begin{equation}
	\varepsilon \lambda \Gamma_0 = \Gamma_0 \left[1 - \exp\left(-\frac{\sigma_c^2}{2\sigma_{c0}^2}\right)\right].
\end{equation}
Solving for $\sigma_c$ yields the tube thickness during reconnection
\begin{equation}\label{eq:sigmac}
	\sigma_c = \sigma_{c0} \sqrt{2 \ln \left( \frac{1}{1 - \varepsilon \lambda} \right)}.
\end{equation}

Substituting \eqref{eq:sigmac} into \eqref{eq:Ni2}, the interaction parameter during reconnection becomes
\begin{equation}\label{Nifinal}
	N_i(\lambda)  = \frac{\lambda N_{i0}}{-2 \ln (1 - \varepsilon \lambda)},
\end{equation}
which is a monotonically increasing function of $\lambda \in (0,1)$. This implies that as reconnection proceeds and the residual flux ratio $\lambda$ decreases, the interaction parameter $N_i$ increases. Consequently, the relative strength of the Lorentz force increases, further suppressing the reconnection process. When the interaction parameter rises beyond a critical threshold that completely inhibits reconnection (as estimated a posteriori from DNS to be $N_{ic} = \mathcal{O}(100)$), the reconnection halts, and the residual flux tubes persist in a partially reconnected state.

However, this variation in the interaction parameter is limited in scope and only becomes significant for moderate initial interaction parameters slightly below the reconnection suppression threshold. From equation~\eqref{Nifinal}, the maximum achievable value of $N_i$ is
\begin{equation}\label{Nimax}
	\max (N_i) = \lim_{\lambda \rightarrow 0} \frac{\lambda N_{i0}}{-2\ln (1 - \varepsilon \lambda)} = \frac{N_{i0}}{2\varepsilon} \approx 1.27 N_{i0},
\end{equation}
which corresponds to only a 27\% increase over the initial value. For very small $N_{i0}$, this increase is insufficient to reach the threshold required to suppress reconnection, thereby allowing full reconnection to proceed. This does not imply, however, that the mechanism driving the growth of the interaction parameter is negligible. On the contrary, it is precisely the presence of such a mechanism that enables vortex tubes and magnetic tubes with moderate initial interaction parameters to retain partial flux over extended periods, thereby facilitating the onset of instabilities and energy cascades.

\begin{figure}
	\centering
	\includegraphics[width=0.6\linewidth]{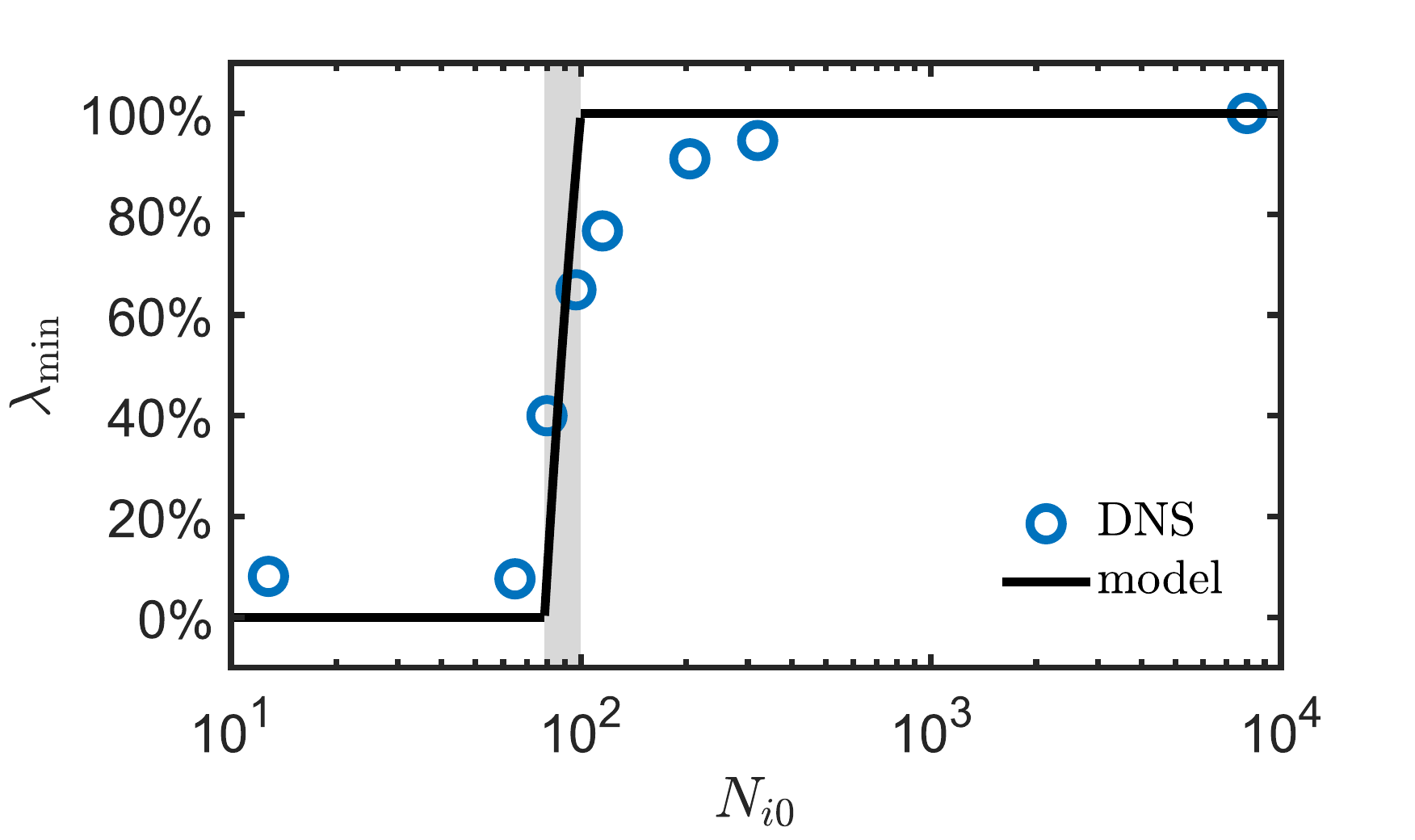}
	\caption{Comparison of DNS (symbols) and model predictions (solid line) for the minimum residual fraction $\lambda_\mathrm{min}$ of $x$-flux at different initial interaction parameters for $\Rey=2000$. Model predictions are calculated from \eqref{Nifinal}.}
	\label{fig:model}
\end{figure}

The flux-transport model predicts both the parameter range in which partial flux transfer occurs and the minimum residual fraction of $x$-flux, $\lambda_\mathrm{min}$, based on the initial conditions, thereby providing an estimate of the degree of reconnection suppression. In the model, complete suppression of flux transfer occurs when the initial interaction parameter satisfies $N_{i0} \geqslant N_{ic}$, while full reconnection takes place when $N_{i0} \leqslant 2\varepsilon N_{ic}$. For the intermediate range $N_{i0} \in (2\varepsilon N_{ic},N_{ic})$, the residual flux fraction for partial reconnection is obtained by solving \eqref{Nifinal} with $N_{ic}$ as the target value. Figure~\ref{fig:model} compares the model predictions with DNS results. Because our model neglects nonlinear effects and considers only a single reconnection event, it underestimates flux transfer in cases where a strong nonlinear cascade develops.

\section{Conclusions}\label{subsec:conclusions}

We have investigated the interplay between vorticity and magnetic fields by imposing magnetic flux tubes of varying strengths on a pair of antiparallel vortex tubes, exploring a wide range of interaction parameters from $N_i = 0$ to $N_i = 32000$. Due to the dual frozen-in conditions for vorticity and magnetic field at high Reynolds numbers, the flow evolution is governed by a competition between vortex dynamics and Lorentz forces. Through DNS, we identify three distinct regimes of vortex–magnetic field interaction: vortex-dominated joint reconnection, instability-triggered cascade, and Lorentz-force-induced vortex disruption.

The limiting cases of pure vortex and pure magnetic flux tubes reveal a fundamental antagonism in their dynamic behavior. Vortex tubes, governed solely by hydrodynamic interactions, promote reconnection through mutual induction and subsequent topological rearrangement, leading to flux transfer across transverse planes. In contrast, magnetic flux tubes, dominated by Lorentz forces and absent induced flow, undergo axial splitting without reconnection, maintaining their original flux alignment in accordance with Alfvén’s theorem.

For small interaction parameters ($N_i \ll \mathcal{O}(10^2)$), classical vortex dynamics dominate. The vortex tubes, carrying frozen-in magnetic flux tubes, undergo a coupled vortex–magnetic reconnection reminiscent of purely hydrodynamic reconnection. Under the frozen-in condition, the magnetic and vortex structures remain tightly coupled. This joint reconnection exhibits both hydrodynamic and magnetohydrodynamic (MHD) characteristics, generating current sheets at reconnection points and leaving remnant magnetic threads. Vortex dynamics drive the stretching of magnetic tubes and threads, acting as a mechanism for dynamo action that converts kinetic energy into magnetic energy. Moreover, increasing the Reynolds number intensifies this process, leading to multi-stage reconnections that enhance the vortex-driven dynamo and promote magnetic energy amplification.

At moderate interaction parameters ($N_i \sim \mathcal{O}(10^2$--$10^3)$), the flow enters a transitional regime where vorticity-induced attraction and magnetic damping are comparable. In this regime, partial reconnection and wave-induced instabilities jointly drive a cascade of multiscale structures. The emergence of secondary filaments causes both kinetic and magnetic energy spectra to approach a $k^{-5/3}$ scaling, indicative of turbulence-like behavior governed by a balance between hydrodynamic and MHD effects. This instability-triggered cascade demonstrates that, even in the absence of a clearly dominant mechanism, energy can be continuously transferred to smaller scales. Such behavior lays the foundation for a coupled vortex–magnetic cascade and a potentially turbulent evolution mediated by hybrid structures.

At high interaction parameters ($N_i \gg \mathcal{O}(10^3)$), Lorentz-force-driven dynamics dominate. The strong Lorentz force rapidly disrupts the coherent vortex cores and suppresses further reconnection, facilitating a rapid conversion of magnetic energy into kinetic energy. The intense magnetic field decouples vortex and magnetic structures, resulting in distinct evolutionary paths: vortex cores are shredded into complex helices, while magnetic flux tubes straighten and remain stable. Despite intense local restructuring, the global circulation remains conserved, and the energy transfer process transitions from transient reconnection bursts to a long-term dissipative decay. This regime marks a shift from instability-driven interactions to magnetically constrained evolution, where turbulence is suppressed and the flow relaxes through slow, magnetically mediated processes.

The presence of magnetic flux tubes has an antagonistic effect on the approach and reconnection of antiparallel vortex tubes. The fluxes transferred across the reconnection region serve as the measure of flux tube reconnection. We observe that increasing the interaction parameter $N_i$ effectively suppresses flux transfer, thereby inhibiting reconnection between the flux tubes. A theoretical analysis shows that, for flux tubes with a Gaussian profile, the reconnection process locally amplifies the interaction parameter, which in turn enhances the magnetic suppression of further reconnection. As a result, even slightly above the reconnection threshold, the system undergoes only partial reconnection and retains a significant fraction of its original magnetic flux over long timescales. This sustained flux preservation promotes the development of instabilities and facilitates the onset of energy cascades. Furthermore, the flux‐transfer model captures the regimes and residual flux fraction of partial reconnection, in reasonable agreement with DNS.

Overall, this study delineates the different pathways of vortex–magnetic structure evolution and energy transfer across a broad range of interaction parameters. These findings provide insight into the structure formation and energy transfer in conducting fluids and have implications for understanding fundamental processes in both astrophysical and industrial plasma systems. Future investigations will focus on extending the parameter space to higher Reynolds numbers and broader magnetic Prandtl number ranges, aiming to explore the dynamics under more realistic astrophysical conditions, such as magnetohydrodynamic processes in the solar convection zone and atmosphere.



\backsection[Acknowledgements]{W. Shen thanks J. Yao and S. Xiong for helpful discussions. Numerical simulations and visualizations were performed on the HPC systems of the Max Planck Institute for Solar System Research and the Max Planck Computing and Data Facility (MPCDF).}

\backsection[Funding]{The authors acknowledge the financial support from the Max Planck Society, the German Research Foundation(DFG) through grants 521319293,540422505 and 550262949.}

\backsection[Declaration of interests]{The authors report no conflict of interest.}

\backsection[Author ORCIDs]{Weiyu Shen https://orcid.org/0000-0003-4385-8835; Rodolfo Ostilla-Mónico https://orcid.org/0000-0001-7049-2432; Xiaojue Zhu https://orcid.org/0000-0002-7878-0655.}


\appendix
\section{Grid convergence analysis}\label{app:grid}
Due to the strong gradients in velocity and magnetic fields, vortex–magnetic joint reconnection imposes stringent requirements on numerical resolution. As the Reynolds number increases, the flow becomes susceptible to inherent symmetry breaking triggered by the Kelvin–Helmholtz (planar jet) instability \citep{Yao2019A,Yao2022Vortex}, making it difficult for the flow evolution to converge with increasing grid resolution due to the influence of numerical errors. In this appendix, a grid sensitivity analysis is conducted for reconnection cases with a low interaction parameter $N_i = 12.8$ at Reynolds numbers $\Rey = 2000$ and $3000$. For each $\Rey$, three different grid resolutions are tested: $N^3 = 128^3$, $256^3$, and $512^3$ for $\Rey = 2000$; and $N^3 = 256^3$, $512^3$, and $1024^3$ for $\Rey = 3000$.

\begin{figure}
	\centering
	\includegraphics[width=\linewidth]{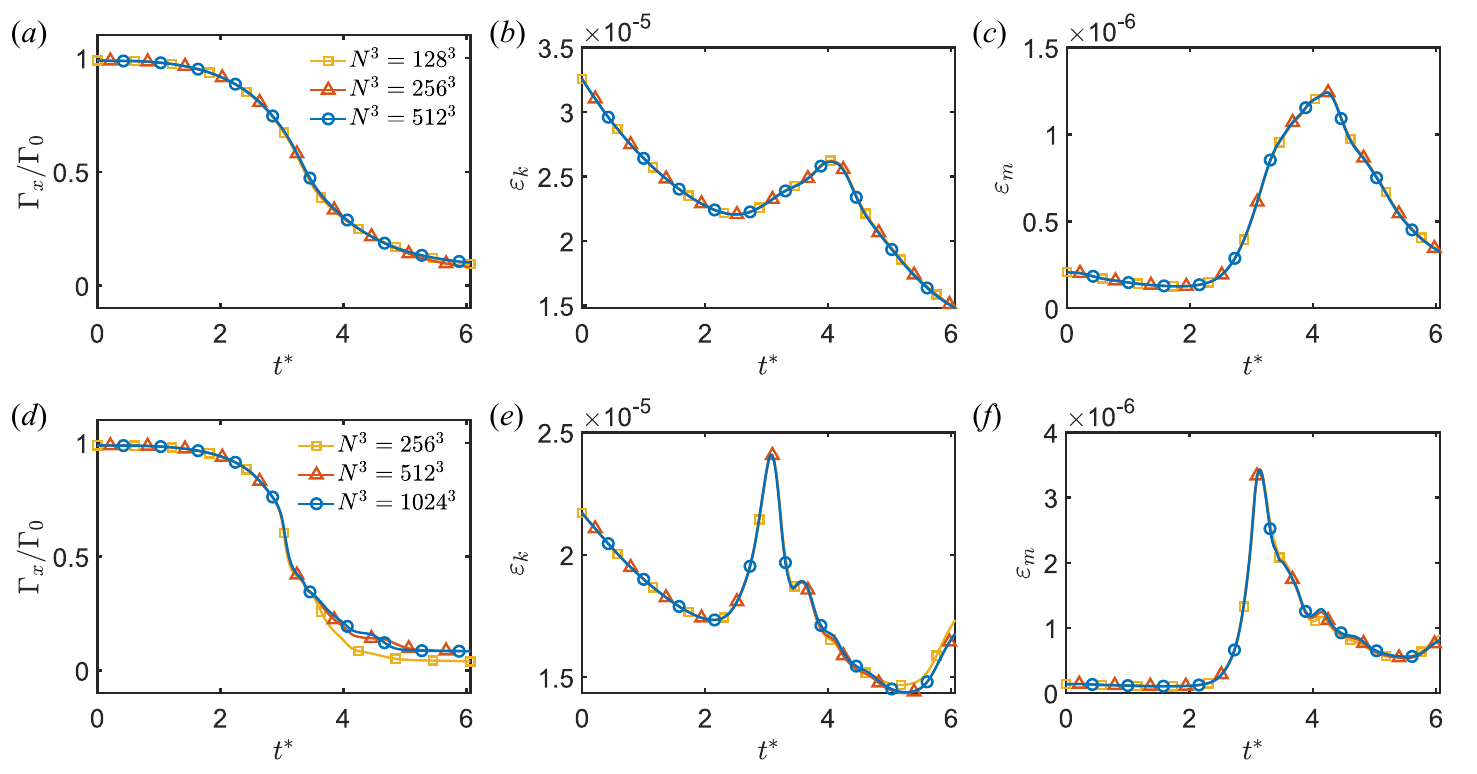}
	\caption{Grid convergence tests for reconnection cases with interaction parameter $N_i = 12.8$: (a,d) circulation transfer, (b,e) kinetic dissipation rate, and (c,f) magnetic dissipation rate at Reynolds numbers (a–c) $\Rey = 2000$ and (d–f) $\Rey = 3000$.}
	\label{fig:grids}
\end{figure}

Figure~\ref{fig:grids} shows the time evolution of the circulation transfer, kinetic dissipation rate, and magnetic dissipation rate for the different grid resolutions. The results from the three grids collapse in figure~\ref{fig:grids}(a–c) for $\Rey = 2000$. The circulation transfer reflects the fidelity of structural evolution. The dissipation rates, as second-order statistics, are highly sensitive to small-scale dynamics. Their collapse indicates that the grid resolutions used in this study are sufficient to resolve the small-scale structures throughout the evolution, including vortex and magnetic features. For $\Rey = 3000$, the flow evolution begins to exhibit effects of symmetry breaking. Nevertheless, all physical quantities still converge to those obtained with the highest resolution $N^3 = 1024^3$ throughout the entire evolution (figure~\ref{fig:grids}(d-f)).

For cases with relatively larger interaction parameters, only partial reconnection occurs, resulting in limited variations in the velocity gradient. Within the parameter range considered in this study, the Lorentz force–driven velocity field remains weaker than the vortex-induced field in the initial state. Consequently, under the same Reynolds number, the grid resolution adequate for lower $N_i$ remains sufficient for higher $N_i$. However, as the magnetic field strength increases, the more rapid flow response demands higher temporal resolution to keep the magnetic CFL number based on $\boldsymbol{b}$ below 0.5 ensuring computational convergence.

\bibliographystyle{jfm}
	
\bibliography{mhd}

\end{document}